\documentclass[twocolumn,aps,pra,superscriptaddress]{revtex4-1}
\usepackage{graphicx}
\usepackage{natbib}
\usepackage{url}
\usepackage{textcase}
\usepackage{bm}
\usepackage{color}
\usepackage{times}
\usepackage{amsmath}

\begin{document}
\title{Rent's rule and extensibility in quantum computing}

\author{D.P. Franke}
\affiliation{QuTech and Kavli Institute of Nanoscience, Delft University of Technology, P.O. Box 5046, 2600 GA Delft, The Netherlands}
\author{J.S. Clarke}
\affiliation{Components Research, Intel Corporation, 2501 NE Century Blvd, Hillsboro, OR 97124, USA}
\author{L.M.K. Vandersypen}
\affiliation{QuTech and Kavli Institute of Nanoscience, Delft University of Technology, P.O. Box 5046, 2600 GA Delft, The Netherlands}
\affiliation{Components Research, Intel Corporation, 2501 NE Century Blvd, Hillsboro, OR 97124, USA}
\author{M. Veldhorst}
\affiliation{QuTech and Kavli Institute of Nanoscience, Delft University of Technology, P.O. Box 5046, 2600 GA Delft, The Netherlands}
\date{\today}

\begin{abstract}
Quantum computing is on the verge of a transition from fundamental research to practical applications. Yet, to make the step to large-scale quantum computation, an extensible qubit system has to be developed. In classical semiconductor technology, this was made possible by the invention of the integrated circuit, which allowed to interconnect large numbers of components without having to solder to each and every one of them. Similarly, we expect that the scaling of interconnections and control lines with the number of qubits will be a central bottleneck in creating large-scale quantum technology. Here, we define the quantum Rent's exponent $p$ to quantify the progress in overcoming this challenge at different levels throughout the quantum computing stack. We further discuss the concept of quantum extensibility as an indicator of a platform's potential to reach the large quantum volume needed for universal quantum computing and review extensibility limits faced by different qubit implementations on the way towards truly large-scale qubit systems.
\end{abstract}

\maketitle

\section{The tyranny of numbers}
One of the most significant advances in the field of quantum computation has been the invention of quantum error correction (QEC) \cite{raussendorf_fault-tolerant_2007, fowler_surface_2012, terhal_quantum_2015}. While quantum bits (qubits) are delicate systems, these algorithms can enable fault-tolerant quantum computation with sophisticated correction codes tolerating error rates of up to 1\% \cite{fowler_surface_2012}. Similar values are already achieved or within reach for experimentally observed qubit infidelities across a range of different platforms \cite{kok_linear_2007, brown_single-qubit-gate_2011, barends_superconducting_2014, waldherr_quantum_2014, dolde_high-fidelity_2014, muhonen_storing_2014, veldhorst_addressable_2014, yoneda_quantum-dot_2017}. However, a trade-off between the tolerated error rates and the number of qubits has to be made. Quantum error correction can lead to an overhead between $10^3$ and $10^4$ physical qubits per logical qubit \cite{fowler_surface_2012, martinis_qubit_2015}, such that millions or even billions of physical qubits will be required for practical applications. To host and control this daunting number of qubits, formidable requirements have to be met by different elements of the system, including interconnects, control electronics and quantum software. It is therefore essential to develop an extensible approach to the hardware and software throughout the full quantum computing stack.

Today, experimental qubit systems make use of one, or even a few, control terminals $T$ per internal component $g$, here the physical qubit \cite{kok_linear_2007, brown_single-qubit-gate_2011, barends_superconducting_2014, waldherr_quantum_2014, dolde_high-fidelity_2014, muhonen_storing_2014, veldhorst_addressable_2014, yoneda_quantum-dot_2017}. This fixed ratio $T/g$ that implies an explosion of the number of terminals with increasing number of components is reminiscent of the late 1950s, where engineers were working with electrical systems containing many component, each requiring soldering to numerous others. Ian Ross, president of Bell labs, stated: `As you built more and more complicated devices, like switching systems, like computers, you got into millions of devices and millions of interconnections. So what should you do?' \cite{gertner_idea_2013}. Jack Morton, vice president of device development at Bell Labs, referred to this situation as `the tyranny of numbers' \cite{morton_technological_1958}. He believed a solution would be to search for devices that could perform multiple tasks, such that the total number of components could be reduced. The real breakthrough was made, among others, by Jack Kilby of Texas Instruments, and Robert Noyce of Fairchild Semiconductor, who improved the Integrated Circuit (IC) to an industrial level. Integrated circuits circumvented the tyranny of numbers and were faster, better, and cheaper.

\begin{figure*}%
	\includegraphics[width=\textwidth]{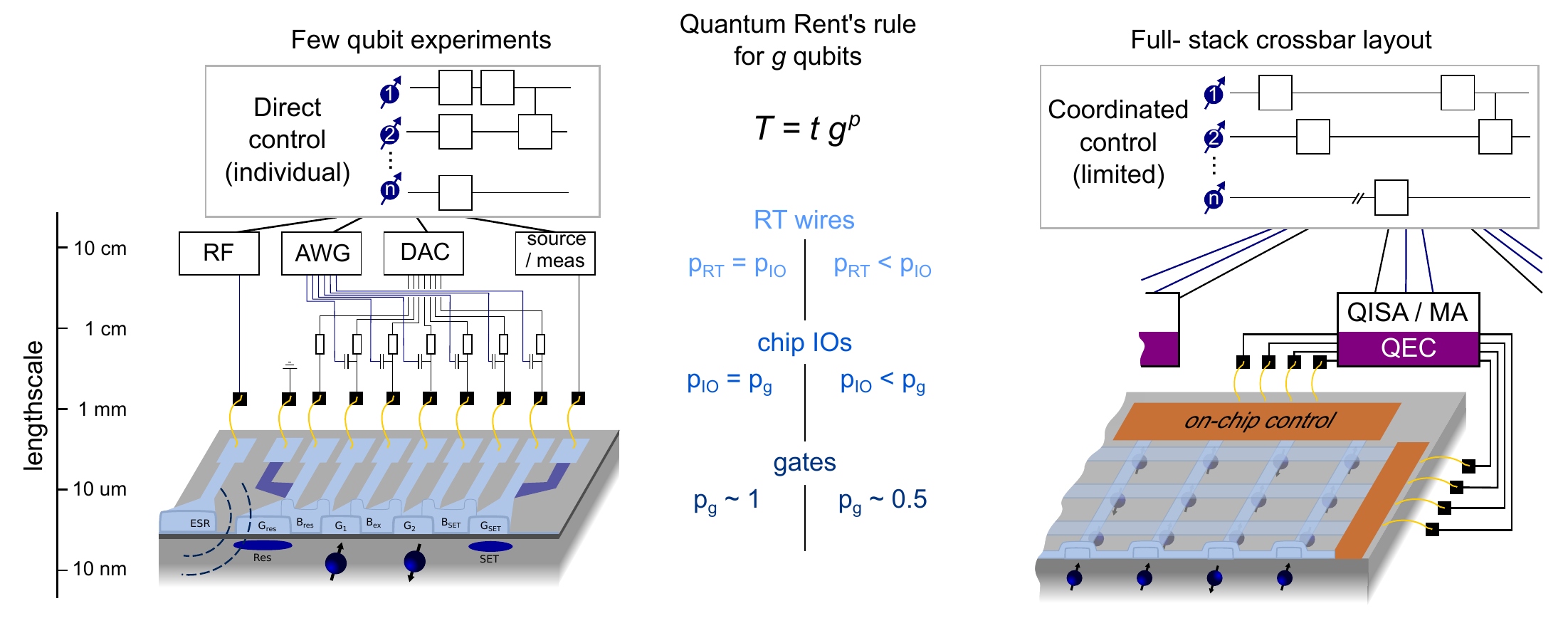}%
	\caption{Comparison of the quantum Rent's component $p$ defined for different layers of a typical few qubit experiment and a possible optimized integration scheme for a spin qubit processor. The exponent $p$ can be improved by solutions at different layers, such as a crossbar gating layout (reducing $p_g$), on-chip routing and multiplexing ($p_{io}$) or cryogenic logic for QEC cycles ($p_{RT}$). As an effect of shared control and the reduced bandwidth per qubit, limitations in the qubit control and the timing of gate sequences will occur as an effect of such optimizations.}
\label{fig:stack}
\end{figure*}

\section{Rent's rule}
An interesting trend between the number of $T$ and $g$ on an IC was observed in the 1960s by E.F Rent, IBM \cite{lanzerotti_microminiature_2005}. Landman and Russo described the correlation using the empirical formula
$$T=tg^p\text{ ,}$$ 
which they called Rent's rule \cite{landman_pin_1971}, and which was later formally justified \cite{christie_interpretation_2000}. Intuitively, $t$ refers to the number of connections required for each internal component $g$. The Rent exponent $p$ accounts for the level of optimization, such that with no optimization $p=1$, while, for example, the X86 series of Intel microprocessors have $p=0.36$ \cite{davis_stochastic_1998}. Guided by the history of classical ICs, we envision that quantum systems, will experience a similar down-scaling in $p$ due to similar motivations. 

To exemplify the corresponding situation in few qubit experiments, a typical measurement setup for quantum dot spin qubits is shown schematically in the left part of Fig.~\ref{fig:stack}. Here, the qubits are controlled by lithographically defined gates and a microwave delivery antenna which fan out to bond pads that are wire-bonded to a chip carrier. Then, the lines are filtered and are wired through the different stages of the dilution refrigerator that keeps the device at its milliKelvin operating temperature. Each line is then individually connected to the outputs of low noise digital analogue converters (DAC) and arbitrary waveform generators (AWG) that are used to control the electronic potential landscape. Adding another qubit to the present device would require an additional two gates and two bonding wires, as well as two additional AWG and DAC channels. This linear scaling is described by an exponent $p=1$ at all levels of the experiment, as indicated in the central column of Fig.~\ref{fig:stack}.

The limitations posed by this scaling law as well as the possible solutions could be very different at different levels of the quantum computer stack. We therefore propose to define several scaling exponents $p$. On the lowest level, $p_g$ describes the number of gates per qubit. Here, typical limitations will be due to geometric restrictions and the limited number of gate layers. For example, at least $\sqrt{g}/2$ gate layers are necessary to directly address $g$ qubits in a 2D array. In close analogy to Rent's rule for IC terminals, $p_{IO}$ describes the number of IO terminals of the chip. Clear limitations are given by the size of these terminals and the space on the chip and, as with classical processors, the number of connections will likely be limited to a few thousand. The third exponent $p_{RT}$ then refers to the number of wires leaving the cryostat. Here, constraints will, for example, be posed by the geometry of the dilution refrigerator and the heat transport trough such wires.
As each of the exponents includes the effect of optimization achieved on lower-lying levels, 
$$ p_g\geq p_{IO}\geq p_{RT} \text{ .}$$

At the current stage, the experimental qubit implementations across all platforms make use of a direct control of each qubit, corresponding to $p_{RT}=1$. This straight-forward implementation at the few-qubits level provides maximum flexibility and control, such that individual adaptations to inhomogeneities are possible. While this concept reduces the demands on the fabrication uniformity, it clearly will not be able to support the large numbers needed for practical error correction. Therefore, schemes of shared control lines have to be developed and implemented at different levels. Several concepts that have been suggested addressing these issues are summarized in the right part of Fig.~\ref{fig:stack}. As proposed in Refs.~\onlinecite{helmer_cavity_2009, hill_surface_2015, veldhorst_silicon_2017, li_crossbar_2017}, two-dimensional arrays and crossbar gating schemes can help to achieve notable improvements in $p_g$, which can be around $p\sim 0.5$ here. The benefit for overcoming the interconnect bottleneck is substantial: with $p\sim 0.5$, one million qubits require not of order one million wires (infeasible) but one thousand wires (feasible). It will be a milestone if such architectures can be realized experimentally.
As a promising way to further reduce the number of IO terminals and hence achieving $p_{IO}<p_g$, cryogenic electronics that can implement on-chip control circuits are pursued \cite{hornibrook_frequency_2014,homulle_cryocmos_2016}. To increase the available cooling power to a level compatible with the dissipation in such circuits, an increase in the qubit operating temperature could be a central step \cite{vandersypen_interfacing_2017, petit_spin_2018}.
Furthermore, local microelectronics and logic circuits used to control error correction cycles or other feedback could be implemented to reduce latencies and trivial communication with room temperature equipment. As a result, $p_{RT}<p_{IO}$. In Fig.~\ref{fig:stack} this is illustrated by the box labeled quantum instruction set architecture (QISA) or microarchitecture (MA).

It is also worth mentioning that such concepts for enhancing the scalability of the wiring will have a direct influence on the way the qubits are operated. While in case of a direct control the highest possible flexibility is maintained, a reduced number of control lines will likely result in an overhead. As suggested in the schematic quantum circuit in Fig.~\ref{fig:stack}, algorithms might have to be restructured since the parallel application of arbitrary pulses to different qubits will be constrained as a result of shared gates. In most cases, this will lead to slower qubit operation giving rise to the question whether this limitation will influence the overall capabilities of a quantum processor. Similarly, the operation of error correction schemes will become more challenging if such a limited control has to be considered. For the example of the crossbar structure proposed in Ref.~\onlinecite{li_crossbar_2017}, it has been shown that surface code operation can indeed be implemented to create a logical qubit with a very low logical error rate even under the limits imposed by shared control \cite{helsen_quantum_2018}. Another particularity of such schemes that remains to be investigated is the influence of correlated error due to the shared control gates. While creating and connecting multiple logical qubits will be a challenging task with shared control, we are optimistic that the advantages of shared control schemes for the extensibility of the qubit devices outweigh the possible restrictions and that they will play a central role in the development of large-scale quantum processors.

\section{Quantum extensibility}
We want to broaden the above discussion to a more general view on the extensibility of proposed and realized qubit systems. 

\begin{figure*}%
	\includegraphics[width=\textwidth]{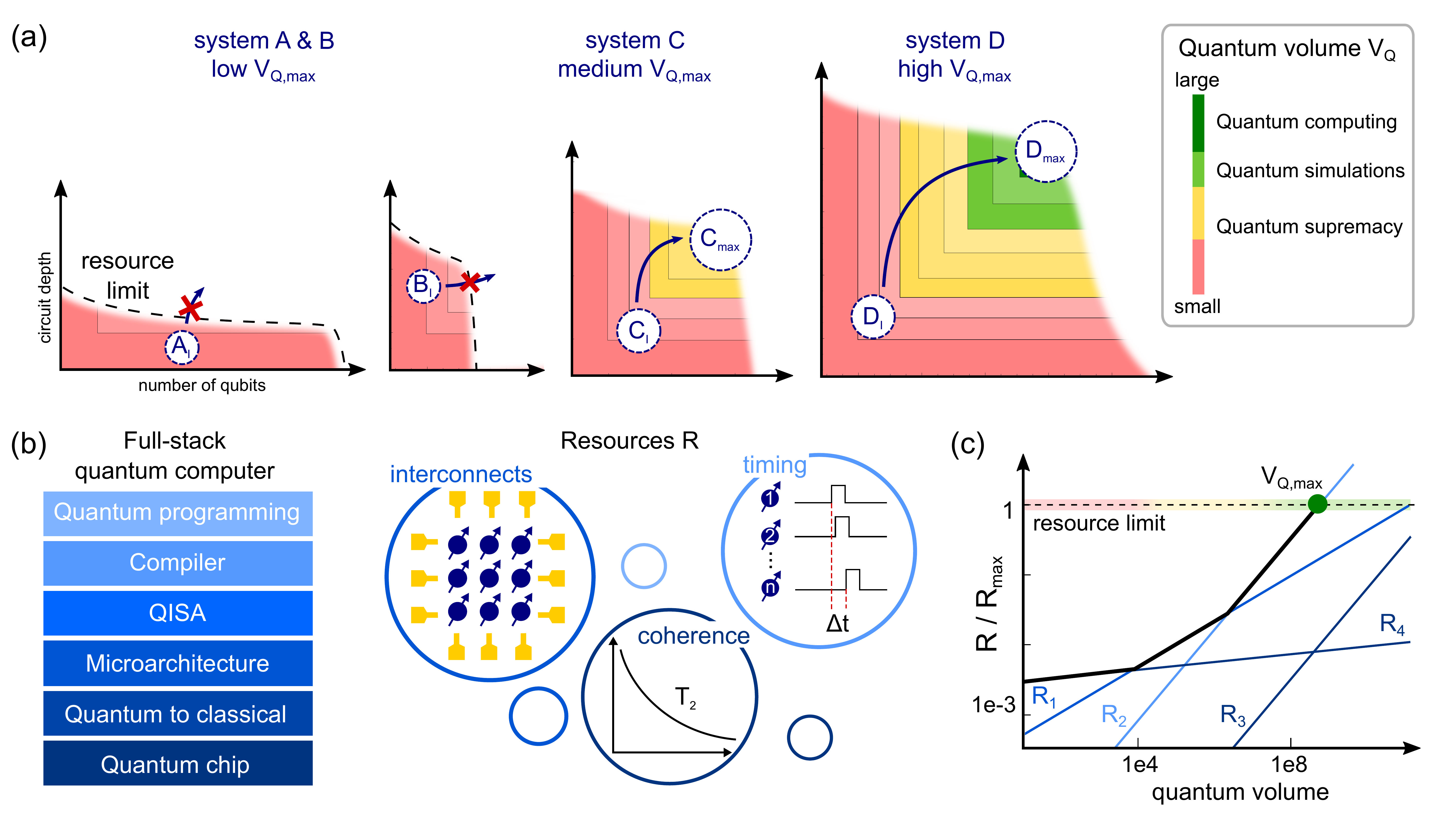}%
	\caption{(a) Quantum volume of the fictional qubit systems A--D. Initial system states are labeled with subscript I, the maximum quantum volume state for system C and D is indicated by $C_\mathrm{max}$ and $D_\mathrm{max}$, respectively. 
	(b) Illustration of the quantum computing stack. Examples for resources that might limit the systems extensibility at different levels are illustrated in circles.
	(c) Extensibility graph for system D, showing the use of four resources as a function of the achievable quantum volume.}
	\label{fig:metrics}
\end{figure*}

\subsection{Quantum volume}
In an attempt to define a metric that can describe the usefulness of a quantum chip by not just the number of qubits but also considering the quality of their implementation, the quantum volume $V_Q$ was introduced by Bishop et al.~\cite{bishop_quantum_2017, moll_quantum_2017}. This metric is a function of the number of qubits $N$ and the circuit depth $d$ of their operations. Here, $d$ is given by the number of operations that can be performed before, on average, an error will occur. It describes to what extent the system can use entanglement and profit from a quantum speed-up \cite{moll_quantum_2017}. The most straight-forward definition of $V_Q$ as the product of $d$ and $N$ already gives a useful metric, but is mostly meaningless when either of the two factors are small. Therefore, the volume is defined as
$$
	V_Q=\min (N,d)^2\text{ .}
$$
In Fig.~\ref{fig:metrics}, color plots of the quantum volume as a function of $N$ and $d$ are shown using logarithmic axes. We discriminate four different regimes as given in the legend of Fig.~\ref{fig:metrics}. The quantum volume of the experimental qubit implementations today is still small and below the threshold where classical computers can still efficiently simulate the quantum system. This is indicated by a red background color. The  yellow background describes quantum systems that might be too complex to be fully simulated by classical computers but are not yet powerful enough to harness the full potential of quantum computing. For these, the term \emph{quantum supremacy} has been coined \cite{preskill_quantum_2012, boixo_characterizing_2018}, which serves as a benchmark in the development of early stage quantum processors. Where exactly this line is drawn is still under active discussion \cite{pednault_breaking_2017, chen_classical_2018}, and specific problems that are designed to be hard on classical computers, but could be solved already by small scale quantum computers have been proposed \cite{harrow_quantum_2017}. From a different viewpoint, this class of devices is also referenced as Noisy Intermediate-Scale Quantum (NISQ) technology \cite{preskill_quantum_2018} and while there might be a limited range of applications, they are mostly considered an intermediate step towards more powerful systems. The two green regions correspond to a quantum volume large enough to allow for relevant quantum simulations \cite{lloyd_universal_1996, cirac_goals_2012} or even fault-tolerant universal quantum computing. When these regions can be reached, it is widely believed that the impact on computing and many other disciplines in science will be revolutionary.

In Fig.~\ref{fig:metrics} (a), the quantum volume is shown for the area in $N$--$d$ space that is expected to be covered for four fictional quantum platforms. Furthermore, the initial state as set by a state-of-the-art system in that platform is shown ($A_I$ etc.), as well as the maximum quantum volume region that can be achieved ($C_\mathrm{max}$ and $D_\mathrm{max}$). In the example of platform A, a rather large number of qubits could be expected to be reached, while their fidelity faces stricter limits and will limit the development of a quantum processor. Platform B, on the other hand, is already in an initial state of better circuit depth, but the number of qubits that can be realized in this approach will restrict the maximum quantum volume. System C will be able to reach beyond quantum supremacy into the NISQ era of quantum applications, but it still lacks the ability to reach a quantum volume large enough for universal quantum computing. Only system D can be expected to reach the green regions of large quantum volume should be considered as a system with real potential for quantum computing.

\subsection{Extensibility}

\newcommand{\vqm}{V_{Q,\mathrm{max}}}
\newcommand{\rmax}{R_\mathrm{max}}
Quantum computer architectures are envisioned as layered control stacks \cite{jones_layered_2012, fu_heterogeneous_2016}, illustrated in Fig. \ref{fig:metrics} (b). 
We expect extensibility limits to occur at all layers of the stack. This ranges from the most fundamental level of the actual physical implementation of qubits that suffers from decoherence limiting the fidelity of qubit operations to more practical limits such as the classical computing power needed to analyze error syndrome measurements in quantum error correction. Other examples include the interconnect bottleneck discussed above, the available cooling power for low temperature operation, space on a chip, or timing issues due to delay in control lines. While these issues are sometimes addressed in a highly speculative way, for example by referring to future developments in fabrication, an honest and preferably quantitative way of giving extensibility limits would be highly beneficial to the field.

We therefore propose that, in addition to estimating the maximum quantum volume $\vqm$, a system is described by its extensibility $X_R$ with respect to a resource $R$. In a generalization of the quantum version of Rent's rule, we assume that the use of most resources can be described by a power law
\begin{align}
	R(V_Q) &= R_I\cdot \left(\frac{V_Q}{V_{Q,I}}\right)^{\frac{1}{X_R}}\text{ ,}\label{eq:ext}
\end{align}
where $V_{Q,I}$ is the initial quantum volume of the system, $R_I$ is the initial use of the resource $R$, and $X_R$ is the extensibility of the system with respect to $R$. In words, the extensibility of a system describes at what expense its quantum volume can be increased. As would be expected, a large extensiblity means that a larger quantum volume can be achieved with only a small increase in resources, while for small $X_R$ only a large increase in $R$ will allow to expand $V_Q$. In the extreme case of exponential scaling of $R$, we define $X_R = 0$. Exponential scaling has in our definition thus zero extensibility, consistent with Feynmann's original view on quantum simulation stating that the number of elements should not explode with the space-time volume of the physical system \cite{feynman_simulating_1982}. For direct comparison and the determination of the most relevant resources, (\ref{eq:ext}) is best written in relation to the resource limit $R_\mathrm{max}$ of $R$, such that
\begin{align*}
	\frac{R(V_Q)}{\rmax}&=r_I\cdot \left(\frac{V_Q}{V_{Q,I}}\right)^{\frac{1}{X_R}}\text{ ,}
\end{align*}
where we have defined $r_I = R_I / \rmax$.

Both the initial use of resources expressed by $r_I$ and the extensibility $X_R$ are relevant to quantify the capability of a system to reach high quantum volumes. This becomes clear from Fig.~\ref{fig:metrics} (c), where the scaling of four fictional resources $R_1, \dots, R_4$ with $V_Q$ is shown. Here, $R_4$ has the highest $r_I$ and is therefore closest to its resource limit at the initial state of low quantum volume. However, because of the high extensibility of the system with respect to $R_4$ ($X_{R4}\gg 1$), this resource will not enforce a relevant limit to the development of the quantum volume. For quantum dot qubits, a similar behavior could be assumed for the effort going into the fabrication of the gate structures. While this effort is already quite high initially (high $r_I$), it is likely that once optimized the fabrication can rather easily be expanded to the creation of large numbers of qubits (high $X$). 
In contrast, a resource with low $r_I$ can still limit the system's development if the corresponding extensibility is low. This is the case for $R_2$ in Fig.~\ref{fig:metrics} (c), which is far from its limit at the initial state, but grows quickly as the system is extended to larger $V_Q$. It is therefore the first to reach the resource limit and will dictate the maximum quantum volume that can be realized. For the resource $R_3$, on the other hand, $r_I$ is so small that $R_3$ is always far from limiting the development, even though $X_{R3}$ is similar to $X_{R2}$. An example for such a resource could be the area on a semiconductor chip occupied by quantum dot qubits. In Fig.~\ref{fig:metrics} (c), the resource that is most strongly limited at any point in the development is emphasized by a thick black line. The slope of this line in the double logarithmic plot defines a local extensibility which could possibly be used to classify the short-term development of a system. The overall extensibility $X$ of the platform, however, is best described by the extensibility with respect to the critically limiting resource, or $X = X_{R2}$ for the system described by Fig.~\ref{fig:metrics} (c).

For most resources, $\rmax$ and hence the relative initial state $r_I$ are not sharply defined or can at least be bent with some effort. Therefore, the extensibility at this maximum quantum volume can give a valuable insight whether a system will be able to profit from such optimizations. This becomes clear, when (\ref{eq:ext}) is solved for $V_{Q,\mathrm{max}}$ such that
\begin{align}
	\vqm = \left(\frac{R_\mathrm{max}}{R_I}\right)^{X_R} V_{Q,I}\text{ ,}
\end{align}
where $R$ refers to the critically limiting resource. For a system with low extensibility, changes to $\rmax$ will barely influence $\vqm$, while for a high $X$, even a small change in $R_\mathrm{max}$ will translate to a large change in the achievable quantum volume $\vqm$. 
It is also clear that even small changes of $X_R$ will have a large influence on $\vqm$. Therefore, studying and understanding the extensibility graph of a system already at an early stage of the development is crucial for judging the promise of a particular quantum technology.

\subsubsection{Quantum chip: Qubit platforms}
Among the broad range of physical systems that are developed for quantum computing \cite{ladd_quantum_2010}, the current prospects of reaching a large quantum volume vary significantly. We will therefore specifically address some of the most prominent approaches to the implementation of qubits.

The qubit platforms that bears most similarities with traditional semiconductor technology is that of semiconductor quantum dot qubits defined in silicon. In fact, the similarity of classical transistors and quantum dot qubits suggests that the fabrication of millions of such qubits will be feasible in the near future. Therefore, the question of the extensibility of the qubit control and the scaling of interconnects as described by the quantum version of Rent's rule are of pressing relevance. Following early proposals of two-dimensional architectures \cite{hollenberg_two-dimensional_2006}, implementations of a shared control based on crossbar designs have been proposed for the related system of donor atoms \cite{hill_surface_2015} and using complementary metal–oxide–semiconductor (CMOS) control elements \cite{veldhorst_silicon_2017}. However, these layouts assume fabrication technology that is far out of reach of that of today’s cutting edge semiconductor technologies. In contrast, a recent proposal where quantum dots are defined using shared gates arrange in a cross-bar architecture can be realized using today's methods \cite{li_crossbar_2017}. Still, as of today, only modest numbers of qubits are operated \cite{watson_programmable_2018} and challenges in fabrication uniformity and control need to be overcome before a large-scale quantum chip becomes feasible.

Having seen remarkable improvements in the qubit properties \cite{devoret_superconducting_2013}, the platform that will likely be the first to reach the NISQ era of quantum applications is that of superconducting qubits. Devices with $\sim 50$ qubits can already be fabricated \cite{noauthor_ibm_2017, noauthor_2018_2018} and concrete steps are taken to achieve the specific goal of reaching the quantum supremacy threshold \cite{neill_blueprint_2018, kelly_preview_2018}. The extensibility beyond the NISQ era is, however, less clear and the implementation of the large number of qubits that will be needed for meaningful quantum computation will pose new challenges. One of these is the physical size of the resonators (on the order of mm), which limits the number qubits that can be fabricated on a wafer. Designs that address the extensibility limit that is posed by the limited number of high frequency connection to the quantum chip have been proposed. This includes a design for a surface code unit cell for both quantum hardware and control signals \cite{versluis_scalable_2017} as well as cavity grids \cite{helmer_cavity_2009}.

Trapped atomic ions are arguably still the most advanced qubit platform today \cite{monz_14-qubit_2011,friis_observation_2018}, but do face serious challenges in their extensibility. In contrast to the other two systems discussed above, they cannot directly be implemented using semiconductor fabrication technology, which could be argued to be the only technology that has been proven capable of the necessary large numbers of components. Therefore, approaches to implement ion traps on semiconductor chips have been realized \cite{stick_ion_2006} and are considered a possible route for scaling \cite{kielpinski_architecture_2002, monroe_scaling_2013, lekitsch_blueprint_2017}. An advantage of the trapped ion approach is that ions in different traps can be entangled with each other via room temperature photonic links in the optical domain\cite{monroe_scaling_2013}, allowing a modular approach that relaxes the interconnect bottleneck. For this to be practical entanglement generation rates have to be increased by order of magnitudes but if this can be realized, the advantage over monolithic quantum circuits is that not all control wires have to interface to a single substrate. Given that optical links allow well-separated modules, even a Rent's exponent $p_g=p_{IO}=1$ may be acceptable, although higher in the stack economic considerations may still enforce $p$ substantially below 1.  The higher operation temperature of ions, which can operate at room temperature and are typically only cooled by liquid nitrogen \cite{lekitsch_blueprint_2017} relaxes other architecture restrictions faced by quantum dot qubits and in particular by superconducting qubits. Nevertheless, the physical size (amounting to about $100\times 100$ m$^2$ for $2\times 10^9$ ions \cite{lekitsch_blueprint_2017}) could make such implementations impractical. Many of the same considerations apply to qubits represented by spins bound to color centers in solids, such as nitrogen-vacancy centers in diamond \cite{wrachtrup_quantum_2001, maurer_room-temperature_2012, doherty_nitrogen-vacancy_2013, taminiau_universal_2014, nemoto_photonic_2014}.

As a final example, there have been proposals for creating quantum circuits based on topological qubits \cite{karzig_scalable_2017}. These could, when they can be realized in the future, profit from certain protected states \cite{freedman_topological_2003}. Depending on the details of the implementation, the number of physical qubits could potentially be reduced compared to QEC concepts in other systems. This may relax requirements on the number of physical qubits, such that a lower extensibility limiting the fabrication of physical qubits in this platform may still allow for practical quantum computation.

\subsubsection{Higher levels}
A central advantage of a systems view of a quantum computer is that higher levels can, to a certain extent, be developed independently of the physical qubit implementation \cite{jones_layered_2012,fu_heterogeneous_2016}. This way, extensibility limits that occur here can already be addressed and specific solutions will likely be beneficial for most platforms. To avoid exploding numbers of off-chip connections, some parts of the control electronics can possibly be integrated with the qubit device. Even the minimum logic signals needed to apply the necessary gates for quantum error correction quickly lead to a bandwidth that will be challenging to realize \cite{jones_layered_2012}, meaning that a basic part of the QEC logic would have to be integrated on-chip. Here, spatial and thermal budgets will play a central role. Other clear resource limits are given by the availability of the classical computing power and memory needed to process error syndrome measurements.
Also, the resources for technically more difficult operations such as microwave control and other fast pulses put constraints on a scalable classical control. The parallel and routed application of electrical signals is therefore necessary and dictates the way quantum gates can be applied to the qubits \cite{hornibrook_cryogenic_2015, asaad_independent_2016, almudever_engineering_2017}.

In addition to such limitations that occur at higher levels of the stack, the efficiency of the QEC is directly related to the extensibility that is achieved at lower levels, such as the quantum chip. There, the use of resources is mostly connected to the number and fidelity of physical qubits. A platform that is capable of a particularly efficient form of QEC can therefore benefit from relaxed restrictions and advances made to these codes will be directly reflected in a higher extensiblity at the lower levels of the stack.
Furthermore, in many cases trade-offs between different resources will have to be made. One example is the concept of shared control discussed above. While the implementation of shared control can significantly improve the extensibility of quantum dot qubits with respect to the number of interconnects and chip terminals, this comes with restrictions to the parallel operation of qubits, which directly influences resources such as the number of operations within the qubit coherence time. 
%
%
%
%
%
%
%
%

\section{Discussion}
Feynman argued in his seminal work \textit{simulating physics with computers}: `The rule of simulation that I would like to have is that the number of computer elements required to simulate a large physical system is only to be proportional to the space-time volume of the physical system. I don't want to have an explosion. That is, if you say I want to explain this much physics, I can do it exactly and I need a certain-sized computer. If doubling the volume of space and time means I'll need an exponentially larger computer, I consider that against the rules (I make up the rules, I'm allowed to do that).' \cite{feynman_simulating_1982}

Here, we have tried to capture this vision by defining Rent exponents across a quantum accelerator stack and by broadening the discussion to include all resources needed for a future quantum computer. The effort required for a platform to reach a certain computing power can be revealed by quantum extensibility graphs.
Where exactly the threshold to useful quantum computing is reached will depend on the development of efficient quantum algorithms and will therefore remain a subject of active research. Similarly, whether the quantum volume as defined above truly reflects the usefulness of a system could depend on the particular use case. In predictions for running Shor's algorithm using quantum error correction on a large scale quantum computer, typically around $N\sim 5000$ logical qubits are used \cite{fowler_surface_2012, suchara_comparing_2013}, suggesting $V_Q\sim 10^7$ as an order of magnitude for relevant computation. In any case, it is clear that the quantum volume $V_Q$ will have to grow by many orders of magnitude to get from the current state to a volume capable of useful quantum computation. It is therefore likely that for the critical resources only an extensibility $X_R>1$ (corresponding to a sublinear scaling) can support such growth.

All platforms will face great challenges in achieving the high extensibility that will allow the development of large-scale quantum computation. For quantum dot qubits in silicon, these challenges bear many parallels to the development of classical integrated circuits and the relation of chip terminals to the number of components can hence be expected to be a central metric for extensibility in this platform. Similar metrics can likely be found in other platforms and should be identified to motivate and focus future research. 
Only if these critical extensibilities can be optimized and non-zero $X_R$ are achieved for all components in a quantum computer, Feynman's vision of harnessing the computational power of an exponentially growing number of quantum states in polynomial time and space will become a reality.

The authors would like to thank W. Lawrie for thorough reading of the manuscript. M.V. acknowledges support by the Netherlands Organization of Scientific Research (NWO) VIDI program.

\bibliography{spinqubit}

\begin{thebibliography}{66}%
\makeatletter
\providecommand \@ifxundefined [1]{%
 \@ifx{#1\undefined}
}%
\providecommand \@ifnum [1]{%
 \ifnum #1\expandafter \@firstoftwo
 \else \expandafter \@secondoftwo
 \fi
}%
\providecommand \@ifx [1]{%
 \ifx #1\expandafter \@firstoftwo
 \else \expandafter \@secondoftwo
 \fi
}%
\providecommand \natexlab [1]{#1}%
\providecommand \enquote  [1]{``#1''}%
\providecommand \bibnamefont  [1]{#1}%
\providecommand \bibfnamefont [1]{#1}%
\providecommand \citenamefont [1]{#1}%
\providecommand \href@noop [0]{\@secondoftwo}%
\providecommand \href [0]{\begingroup \@sanitize@url \@href}%
\providecommand \@href[1]{\@@startlink{#1}\@@href}%
\providecommand \@@href[1]{\endgroup#1\@@endlink}%
\providecommand \@sanitize@url [0]{\catcode `\\12\catcode `\$12\catcode
  `\&12\catcode `\#12\catcode `\^12\catcode `\_12\catcode `\%12\relax}%
\providecommand \@@startlink[1]{}%
\providecommand \@@endlink[0]{}%
\providecommand \url  [0]{\begingroup\@sanitize@url \@url }%
\providecommand \@url [1]{\endgroup\@href {#1}{\urlprefix }}%
\providecommand \urlprefix  [0]{URL }%
\providecommand \Eprint [0]{\href }%
\providecommand \doibase [0]{http://dx.doi.org/}%
\providecommand \selectlanguage [0]{\@gobble}%
\providecommand \bibinfo  [0]{\@secondoftwo}%
\providecommand \bibfield  [0]{\@secondoftwo}%
\providecommand \translation [1]{[#1]}%
\providecommand \BibitemOpen [0]{}%
\providecommand \bibitemStop [0]{}%
\providecommand \bibitemNoStop [0]{.\EOS\space}%
\providecommand \EOS [0]{\spacefactor3000\relax}%
\providecommand \BibitemShut  [1]{\csname bibitem#1\endcsname}%
\let\auto@bib@innerbib\@empty
\bibitem [{\citenamefont {Raussendorf}\ and\ \citenamefont
  {Harrington}(2007)}]{raussendorf_fault-tolerant_2007}%
  \BibitemOpen
  \bibfield  {author} {\bibinfo {author} {\bibfnamefont {R.}~\bibnamefont
  {Raussendorf}}\ and\ \bibinfo {author} {\bibfnamefont {J.}~\bibnamefont
  {Harrington}},\ }\href {\doibase 10.1103/PhysRevLett.98.190504} {\bibfield
  {journal} {\bibinfo  {journal} {Phys. Rev. Lett.}\ }\textbf {\bibinfo
  {volume} {98}},\ \bibinfo {pages} {190504} (\bibinfo {year}
  {2007})}\BibitemShut {NoStop}%
\bibitem [{\citenamefont {Fowler}\ \emph {et~al.}(2012)\citenamefont {Fowler},
  \citenamefont {Mariantoni}, \citenamefont {Martinis},\ and\ \citenamefont
  {Cleland}}]{fowler_surface_2012}%
  \BibitemOpen
  \bibfield  {author} {\bibinfo {author} {\bibfnamefont {A.~G.}\ \bibnamefont
  {Fowler}}, \bibinfo {author} {\bibfnamefont {M.}~\bibnamefont {Mariantoni}},
  \bibinfo {author} {\bibfnamefont {J.~M.}\ \bibnamefont {Martinis}}, \ and\
  \bibinfo {author} {\bibfnamefont {A.~N.}\ \bibnamefont {Cleland}},\ }\href
  {\doibase 10.1103/PhysRevA.86.032324} {\bibfield  {journal} {\bibinfo
  {journal} {Phys. Rev. A}\ }\textbf {\bibinfo {volume} {86}},\ \bibinfo
  {pages} {032324} (\bibinfo {year} {2012})}\BibitemShut {NoStop}%
\bibitem [{\citenamefont {Terhal}(2015)}]{terhal_quantum_2015}%
  \BibitemOpen
  \bibfield  {author} {\bibinfo {author} {\bibfnamefont {B.~M.}\ \bibnamefont
  {Terhal}},\ }\href {\doibase 10.1103/RevModPhys.87.307} {\bibfield  {journal}
  {\bibinfo  {journal} {Rev. Mod. Phys.}\ }\textbf {\bibinfo {volume} {87}},\
  \bibinfo {pages} {307} (\bibinfo {year} {2015})}\BibitemShut {NoStop}%
\bibitem [{\citenamefont {Kok}\ \emph {et~al.}(2007)\citenamefont {Kok},
  \citenamefont {Munro}, \citenamefont {Nemoto}, \citenamefont {Ralph},
  \citenamefont {Dowling},\ and\ \citenamefont {Milburn}}]{kok_linear_2007}%
  \BibitemOpen
  \bibfield  {author} {\bibinfo {author} {\bibfnamefont {P.}~\bibnamefont
  {Kok}}, \bibinfo {author} {\bibfnamefont {W.~J.}\ \bibnamefont {Munro}},
  \bibinfo {author} {\bibfnamefont {K.}~\bibnamefont {Nemoto}}, \bibinfo
  {author} {\bibfnamefont {T.~C.}\ \bibnamefont {Ralph}}, \bibinfo {author}
  {\bibfnamefont {J.~P.}\ \bibnamefont {Dowling}}, \ and\ \bibinfo {author}
  {\bibfnamefont {G.~J.}\ \bibnamefont {Milburn}},\ }\href {\doibase
  10.1103/RevModPhys.79.135} {\bibfield  {journal} {\bibinfo  {journal} {Rev.
  Mod. Phys.}\ }\textbf {\bibinfo {volume} {79}},\ \bibinfo {pages} {135}
  (\bibinfo {year} {2007})}\BibitemShut {NoStop}%
\bibitem [{\citenamefont {Brown}\ \emph {et~al.}(2011)\citenamefont {Brown},
  \citenamefont {Wilson}, \citenamefont {Colombe}, \citenamefont {Ospelkaus},
  \citenamefont {Meier}, \citenamefont {Knill}, \citenamefont {Leibfried},\
  and\ \citenamefont {Wineland}}]{brown_single-qubit-gate_2011}%
  \BibitemOpen
  \bibfield  {author} {\bibinfo {author} {\bibfnamefont {K.~R.}\ \bibnamefont
  {Brown}}, \bibinfo {author} {\bibfnamefont {A.~C.}\ \bibnamefont {Wilson}},
  \bibinfo {author} {\bibfnamefont {Y.}~\bibnamefont {Colombe}}, \bibinfo
  {author} {\bibfnamefont {C.}~\bibnamefont {Ospelkaus}}, \bibinfo {author}
  {\bibfnamefont {A.~M.}\ \bibnamefont {Meier}}, \bibinfo {author}
  {\bibfnamefont {E.}~\bibnamefont {Knill}}, \bibinfo {author} {\bibfnamefont
  {D.}~\bibnamefont {Leibfried}}, \ and\ \bibinfo {author} {\bibfnamefont
  {D.~J.}\ \bibnamefont {Wineland}},\ }\href {\doibase
  10.1103/PhysRevA.84.030303} {\bibfield  {journal} {\bibinfo  {journal} {Phys.
  Rev. A}\ }\textbf {\bibinfo {volume} {84}},\ \bibinfo {pages} {030303}
  (\bibinfo {year} {2011})}\BibitemShut {NoStop}%
\bibitem [{\citenamefont {Barends}\ \emph {et~al.}(2014)\citenamefont
  {Barends}, \citenamefont {Kelly}, \citenamefont {Megrant}, \citenamefont
  {Veitia}, \citenamefont {Sank}, \citenamefont {Jeffrey}, \citenamefont
  {White}, \citenamefont {Mutus}, \citenamefont {Fowler}, \citenamefont
  {Campbell}, \citenamefont {Chen}, \citenamefont {Chen}, \citenamefont
  {Chiaro}, \citenamefont {Dunsworth}, \citenamefont {Neill}, \citenamefont
  {O’Malley}, \citenamefont {Roushan}, \citenamefont {Vainsencher},
  \citenamefont {Wenner}, \citenamefont {Korotkov}, \citenamefont {Cleland},\
  and\ \citenamefont {Martinis}}]{barends_superconducting_2014}%
  \BibitemOpen
  \bibfield  {author} {\bibinfo {author} {\bibfnamefont {R.}~\bibnamefont
  {Barends}}, \bibinfo {author} {\bibfnamefont {J.}~\bibnamefont {Kelly}},
  \bibinfo {author} {\bibfnamefont {A.}~\bibnamefont {Megrant}}, \bibinfo
  {author} {\bibfnamefont {A.}~\bibnamefont {Veitia}}, \bibinfo {author}
  {\bibfnamefont {D.}~\bibnamefont {Sank}}, \bibinfo {author} {\bibfnamefont
  {E.}~\bibnamefont {Jeffrey}}, \bibinfo {author} {\bibfnamefont {T.~C.}\
  \bibnamefont {White}}, \bibinfo {author} {\bibfnamefont {J.}~\bibnamefont
  {Mutus}}, \bibinfo {author} {\bibfnamefont {A.~G.}\ \bibnamefont {Fowler}},
  \bibinfo {author} {\bibfnamefont {B.}~\bibnamefont {Campbell}}, \bibinfo
  {author} {\bibfnamefont {Y.}~\bibnamefont {Chen}}, \bibinfo {author}
  {\bibfnamefont {Z.}~\bibnamefont {Chen}}, \bibinfo {author} {\bibfnamefont
  {B.}~\bibnamefont {Chiaro}}, \bibinfo {author} {\bibfnamefont
  {A.}~\bibnamefont {Dunsworth}}, \bibinfo {author} {\bibfnamefont
  {C.}~\bibnamefont {Neill}}, \bibinfo {author} {\bibfnamefont
  {P.}~\bibnamefont {O’Malley}}, \bibinfo {author} {\bibfnamefont
  {P.}~\bibnamefont {Roushan}}, \bibinfo {author} {\bibfnamefont
  {A.}~\bibnamefont {Vainsencher}}, \bibinfo {author} {\bibfnamefont
  {J.}~\bibnamefont {Wenner}}, \bibinfo {author} {\bibfnamefont {A.~N.}\
  \bibnamefont {Korotkov}}, \bibinfo {author} {\bibfnamefont {A.~N.}\
  \bibnamefont {Cleland}}, \ and\ \bibinfo {author} {\bibfnamefont {J.~M.}\
  \bibnamefont {Martinis}},\ }\href {\doibase 10.1038/nature13171} {\bibfield
  {journal} {\bibinfo  {journal} {Nature}\ }\textbf {\bibinfo {volume} {508}},\
  \bibinfo {pages} {500} (\bibinfo {year} {2014})}\BibitemShut {NoStop}%
\bibitem [{\citenamefont {Waldherr}\ \emph {et~al.}(2014)\citenamefont
  {Waldherr}, \citenamefont {Wang}, \citenamefont {Zaiser}, \citenamefont
  {Jamali}, \citenamefont {Schulte-Herbrüggen}, \citenamefont {Abe},
  \citenamefont {Ohshima}, \citenamefont {Isoya}, \citenamefont {Du},
  \citenamefont {Neumann},\ and\ \citenamefont
  {Wrachtrup}}]{waldherr_quantum_2014}%
  \BibitemOpen
  \bibfield  {author} {\bibinfo {author} {\bibfnamefont {G.}~\bibnamefont
  {Waldherr}}, \bibinfo {author} {\bibfnamefont {Y.}~\bibnamefont {Wang}},
  \bibinfo {author} {\bibfnamefont {S.}~\bibnamefont {Zaiser}}, \bibinfo
  {author} {\bibfnamefont {M.}~\bibnamefont {Jamali}}, \bibinfo {author}
  {\bibfnamefont {T.}~\bibnamefont {Schulte-Herbrüggen}}, \bibinfo {author}
  {\bibfnamefont {H.}~\bibnamefont {Abe}}, \bibinfo {author} {\bibfnamefont
  {T.}~\bibnamefont {Ohshima}}, \bibinfo {author} {\bibfnamefont
  {J.}~\bibnamefont {Isoya}}, \bibinfo {author} {\bibfnamefont {J.~F.}\
  \bibnamefont {Du}}, \bibinfo {author} {\bibfnamefont {P.}~\bibnamefont
  {Neumann}}, \ and\ \bibinfo {author} {\bibfnamefont {J.}~\bibnamefont
  {Wrachtrup}},\ }\href {\doibase 10.1038/nature12919} {\bibfield  {journal}
  {\bibinfo  {journal} {Nature}\ }\textbf {\bibinfo {volume} {506}},\ \bibinfo
  {pages} {204} (\bibinfo {year} {2014})}\BibitemShut {NoStop}%
\bibitem [{\citenamefont {Dolde}\ \emph {et~al.}(2014)\citenamefont {Dolde},
  \citenamefont {Bergholm}, \citenamefont {Wang}, \citenamefont {Jakobi},
  \citenamefont {Naydenov}, \citenamefont {Pezzagna}, \citenamefont {Meijer},
  \citenamefont {Jelezko}, \citenamefont {Neumann}, \citenamefont
  {Schulte-Herbrueggen}, \citenamefont {Biamonte},\ and\ \citenamefont
  {Wrachtrup}}]{dolde_high-fidelity_2014}%
  \BibitemOpen
  \bibfield  {author} {\bibinfo {author} {\bibfnamefont {F.}~\bibnamefont
  {Dolde}}, \bibinfo {author} {\bibfnamefont {V.}~\bibnamefont {Bergholm}},
  \bibinfo {author} {\bibfnamefont {Y.}~\bibnamefont {Wang}}, \bibinfo {author}
  {\bibfnamefont {I.}~\bibnamefont {Jakobi}}, \bibinfo {author} {\bibfnamefont
  {B.}~\bibnamefont {Naydenov}}, \bibinfo {author} {\bibfnamefont
  {S.}~\bibnamefont {Pezzagna}}, \bibinfo {author} {\bibfnamefont
  {J.}~\bibnamefont {Meijer}}, \bibinfo {author} {\bibfnamefont
  {F.}~\bibnamefont {Jelezko}}, \bibinfo {author} {\bibfnamefont
  {P.}~\bibnamefont {Neumann}}, \bibinfo {author} {\bibfnamefont
  {T.}~\bibnamefont {Schulte-Herbrueggen}}, \bibinfo {author} {\bibfnamefont
  {J.}~\bibnamefont {Biamonte}}, \ and\ \bibinfo {author} {\bibfnamefont
  {J.}~\bibnamefont {Wrachtrup}},\ }\href {\doibase 10.1038/ncomms4371}
  {\bibfield  {journal} {\bibinfo  {journal} {Nat. Commun.}\ }\textbf {\bibinfo
  {volume} {5}},\ \bibinfo {pages} {3371} (\bibinfo {year} {2014})}\BibitemShut
  {NoStop}%
\bibitem [{\citenamefont {Muhonen}\ \emph {et~al.}(2014)\citenamefont
  {Muhonen}, \citenamefont {Dehollain}, \citenamefont {Laucht}, \citenamefont
  {Hudson}, \citenamefont {Kalra}, \citenamefont {Sekiguchi}, \citenamefont
  {Itoh}, \citenamefont {Jamieson}, \citenamefont {McCallum}, \citenamefont
  {Dzurak},\ and\ \citenamefont {Morello}}]{muhonen_storing_2014}%
  \BibitemOpen
  \bibfield  {author} {\bibinfo {author} {\bibfnamefont {J.~T.}\ \bibnamefont
  {Muhonen}}, \bibinfo {author} {\bibfnamefont {J.~P.}\ \bibnamefont
  {Dehollain}}, \bibinfo {author} {\bibfnamefont {A.}~\bibnamefont {Laucht}},
  \bibinfo {author} {\bibfnamefont {F.~E.}\ \bibnamefont {Hudson}}, \bibinfo
  {author} {\bibfnamefont {R.}~\bibnamefont {Kalra}}, \bibinfo {author}
  {\bibfnamefont {T.}~\bibnamefont {Sekiguchi}}, \bibinfo {author}
  {\bibfnamefont {K.~M.}\ \bibnamefont {Itoh}}, \bibinfo {author}
  {\bibfnamefont {D.~N.}\ \bibnamefont {Jamieson}}, \bibinfo {author}
  {\bibfnamefont {J.~C.}\ \bibnamefont {McCallum}}, \bibinfo {author}
  {\bibfnamefont {A.~S.}\ \bibnamefont {Dzurak}}, \ and\ \bibinfo {author}
  {\bibfnamefont {A.}~\bibnamefont {Morello}},\ }\href {\doibase
  10.1038/nnano.2014.211} {\bibfield  {journal} {\bibinfo  {journal} {Nat.
  Nanotech.}\ }\textbf {\bibinfo {volume} {9}},\ \bibinfo {pages} {986}
  (\bibinfo {year} {2014})}\BibitemShut {NoStop}%
\bibitem [{\citenamefont {Veldhorst}\ \emph {et~al.}(2014)\citenamefont
  {Veldhorst}, \citenamefont {Hwang}, \citenamefont {Yang}, \citenamefont
  {Leenstra}, \citenamefont {Ronde}, \citenamefont {Dehollain}, \citenamefont
  {Muhonen}, \citenamefont {Hudson}, \citenamefont {Itoh}, \citenamefont
  {Morello},\ and\ \citenamefont {Dzurak}}]{veldhorst_addressable_2014}%
  \BibitemOpen
  \bibfield  {author} {\bibinfo {author} {\bibfnamefont {M.}~\bibnamefont
  {Veldhorst}}, \bibinfo {author} {\bibfnamefont {J.~C.~C.}\ \bibnamefont
  {Hwang}}, \bibinfo {author} {\bibfnamefont {C.~H.}\ \bibnamefont {Yang}},
  \bibinfo {author} {\bibfnamefont {A.~W.}\ \bibnamefont {Leenstra}}, \bibinfo
  {author} {\bibfnamefont {B.~d.}\ \bibnamefont {Ronde}}, \bibinfo {author}
  {\bibfnamefont {J.~P.}\ \bibnamefont {Dehollain}}, \bibinfo {author}
  {\bibfnamefont {J.~T.}\ \bibnamefont {Muhonen}}, \bibinfo {author}
  {\bibfnamefont {F.~E.}\ \bibnamefont {Hudson}}, \bibinfo {author}
  {\bibfnamefont {K.~M.}\ \bibnamefont {Itoh}}, \bibinfo {author}
  {\bibfnamefont {A.}~\bibnamefont {Morello}}, \ and\ \bibinfo {author}
  {\bibfnamefont {A.~S.}\ \bibnamefont {Dzurak}},\ }\href {\doibase
  10.1038/nnano.2014.216} {\bibfield  {journal} {\bibinfo  {journal} {Nat.
  Nanotech.}\ }\textbf {\bibinfo {volume} {9}},\ \bibinfo {pages} {981}
  (\bibinfo {year} {2014})}\BibitemShut {NoStop}%
\bibitem [{\citenamefont {Yoneda}\ \emph {et~al.}(2017)\citenamefont {Yoneda},
  \citenamefont {Takeda}, \citenamefont {Otsuka}, \citenamefont {Nakajima},
  \citenamefont {Delbecq}, \citenamefont {Allison}, \citenamefont {Honda},
  \citenamefont {Kodera}, \citenamefont {Oda}, \citenamefont {Hoshi},
  \citenamefont {Usami}, \citenamefont {Itoh},\ and\ \citenamefont
  {Tarucha}}]{yoneda_quantum-dot_2017}%
  \BibitemOpen
  \bibfield  {author} {\bibinfo {author} {\bibfnamefont {J.}~\bibnamefont
  {Yoneda}}, \bibinfo {author} {\bibfnamefont {K.}~\bibnamefont {Takeda}},
  \bibinfo {author} {\bibfnamefont {T.}~\bibnamefont {Otsuka}}, \bibinfo
  {author} {\bibfnamefont {T.}~\bibnamefont {Nakajima}}, \bibinfo {author}
  {\bibfnamefont {M.~R.}\ \bibnamefont {Delbecq}}, \bibinfo {author}
  {\bibfnamefont {G.}~\bibnamefont {Allison}}, \bibinfo {author} {\bibfnamefont
  {T.}~\bibnamefont {Honda}}, \bibinfo {author} {\bibfnamefont
  {T.}~\bibnamefont {Kodera}}, \bibinfo {author} {\bibfnamefont
  {S.}~\bibnamefont {Oda}}, \bibinfo {author} {\bibfnamefont {Y.}~\bibnamefont
  {Hoshi}}, \bibinfo {author} {\bibfnamefont {N.}~\bibnamefont {Usami}},
  \bibinfo {author} {\bibfnamefont {K.~M.}\ \bibnamefont {Itoh}}, \ and\
  \bibinfo {author} {\bibfnamefont {S.}~\bibnamefont {Tarucha}},\ }\href
  {\doibase 10.1038/s41565-017-0014-x} {\bibfield  {journal} {\bibinfo
  {journal} {Nat. Nanotech.}\ ,\ \bibinfo {pages} {102}} (\bibinfo {year}
  {2017})}\BibitemShut {NoStop}%
\bibitem [{\citenamefont {Martinis}(2015)}]{martinis_qubit_2015}%
  \BibitemOpen
  \bibfield  {author} {\bibinfo {author} {\bibfnamefont {J.~M.}\ \bibnamefont
  {Martinis}},\ }\href {\doibase 10.1038/npjqi.2015.5} {\bibfield  {journal}
  {\bibinfo  {journal} {npj Quantum Information}\ }\textbf {\bibinfo {volume}
  {1}},\ \bibinfo {pages} {15005} (\bibinfo {year} {2015})}\BibitemShut
  {NoStop}%
\bibitem [{\citenamefont {Gertner}(2013)}]{gertner_idea_2013}%
  \BibitemOpen
  \bibfield  {author} {\bibinfo {author} {\bibfnamefont {J.}~\bibnamefont
  {Gertner}},\ }\href@noop {} {\emph {\bibinfo {title} {The idea factory:
  {Bell} {Labs} and the great age of {American} innovation}}}\ (\bibinfo
  {publisher} {Penguin Press},\ \bibinfo {year} {2013})\BibitemShut {NoStop}%
\bibitem [{\citenamefont {Morton}\ and\ \citenamefont
  {Pietenpol}(1958)}]{morton_technological_1958}%
  \BibitemOpen
  \bibfield  {author} {\bibinfo {author} {\bibfnamefont {J.~A.}\ \bibnamefont
  {Morton}}\ and\ \bibinfo {author} {\bibfnamefont {W.~J.}\ \bibnamefont
  {Pietenpol}},\ }\href {\doibase 10.1109/JRPROC.1958.286834} {\bibfield
  {journal} {\bibinfo  {journal} {Proceedings of the IRE}\ }\textbf {\bibinfo
  {volume} {46}},\ \bibinfo {pages} {955} (\bibinfo {year} {1958})}\BibitemShut
  {NoStop}%
\bibitem [{\citenamefont {Lanzerotti}\ \emph {et~al.}(2005)\citenamefont
  {Lanzerotti}, \citenamefont {Fiorenza},\ and\ \citenamefont
  {Rand}}]{lanzerotti_microminiature_2005}%
  \BibitemOpen
  \bibfield  {author} {\bibinfo {author} {\bibfnamefont {M.~Y.}\ \bibnamefont
  {Lanzerotti}}, \bibinfo {author} {\bibfnamefont {G.}~\bibnamefont
  {Fiorenza}}, \ and\ \bibinfo {author} {\bibfnamefont {R.~A.}\ \bibnamefont
  {Rand}},\ }\href {\doibase 10.1147/rd.494.0777} {\bibfield  {journal}
  {\bibinfo  {journal} {IBM Journal of Research and Development}\ }\textbf
  {\bibinfo {volume} {49}},\ \bibinfo {pages} {777} (\bibinfo {year}
  {2005})}\BibitemShut {NoStop}%
\bibitem [{\citenamefont {Landman}\ and\ \citenamefont
  {Russo}(1971)}]{landman_pin_1971}%
  \BibitemOpen
  \bibfield  {author} {\bibinfo {author} {\bibfnamefont {B.~S.}\ \bibnamefont
  {Landman}}\ and\ \bibinfo {author} {\bibfnamefont {R.~L.}\ \bibnamefont
  {Russo}},\ }\href {\doibase 10.1109/T-C.1971.223159} {\bibfield  {journal}
  {\bibinfo  {journal} {IEEE Trans. Comput.}\ }\textbf {\bibinfo {volume}
  {20}},\ \bibinfo {pages} {1469} (\bibinfo {year} {1971})}\BibitemShut
  {NoStop}%
\bibitem [{\citenamefont {Christie}\ and\ \citenamefont
  {Stroobandt}(2000)}]{christie_interpretation_2000}%
  \BibitemOpen
  \bibfield  {author} {\bibinfo {author} {\bibfnamefont {P.}~\bibnamefont
  {Christie}}\ and\ \bibinfo {author} {\bibfnamefont {D.}~\bibnamefont
  {Stroobandt}},\ }\href {\doibase 10.1109/92.902258} {\bibfield  {journal}
  {\bibinfo  {journal} {IEEE Transactions on Very Large Scale Integration
  Systems}\ }\textbf {\bibinfo {volume} {8}},\ \bibinfo {pages} {639} (\bibinfo
  {year} {2000})}\BibitemShut {NoStop}%
\bibitem [{\citenamefont {Davis}\ \emph {et~al.}(1998)\citenamefont {Davis},
  \citenamefont {De},\ and\ \citenamefont {Meindl}}]{davis_stochastic_1998}%
  \BibitemOpen
  \bibfield  {author} {\bibinfo {author} {\bibfnamefont {J.~A.}\ \bibnamefont
  {Davis}}, \bibinfo {author} {\bibfnamefont {V.~K.}\ \bibnamefont {De}}, \
  and\ \bibinfo {author} {\bibfnamefont {J.~D.}\ \bibnamefont {Meindl}},\
  }\href {\doibase 10.1109/16.661219} {\bibfield  {journal} {\bibinfo
  {journal} {IEEE Transactions on Electron Devices}\ }\textbf {\bibinfo
  {volume} {45}},\ \bibinfo {pages} {580} (\bibinfo {year} {1998})}\BibitemShut
  {NoStop}%
\bibitem [{\citenamefont {Helmer}\ \emph {et~al.}(2009)\citenamefont {Helmer},
  \citenamefont {Mariantoni}, \citenamefont {Fowler}, \citenamefont {Delft},
  \citenamefont {Solano},\ and\ \citenamefont
  {Marquardt}}]{helmer_cavity_2009}%
  \BibitemOpen
  \bibfield  {author} {\bibinfo {author} {\bibfnamefont {F.}~\bibnamefont
  {Helmer}}, \bibinfo {author} {\bibfnamefont {M.}~\bibnamefont {Mariantoni}},
  \bibinfo {author} {\bibfnamefont {A.~G.}\ \bibnamefont {Fowler}}, \bibinfo
  {author} {\bibfnamefont {J.~v.}\ \bibnamefont {Delft}}, \bibinfo {author}
  {\bibfnamefont {E.}~\bibnamefont {Solano}}, \ and\ \bibinfo {author}
  {\bibfnamefont {F.}~\bibnamefont {Marquardt}},\ }\href {\doibase
  10.1209/0295-5075/85/50007} {\bibfield  {journal} {\bibinfo  {journal} {EPL}\
  }\textbf {\bibinfo {volume} {85}},\ \bibinfo {pages} {50007} (\bibinfo {year}
  {2009})}\BibitemShut {NoStop}%
\bibitem [{\citenamefont {Hill}\ \emph {et~al.}(2015)\citenamefont {Hill},
  \citenamefont {Peretz}, \citenamefont {Hile}, \citenamefont {House},
  \citenamefont {Fuechsle}, \citenamefont {Rogge}, \citenamefont {Simmons},\
  and\ \citenamefont {Hollenberg}}]{hill_surface_2015}%
  \BibitemOpen
  \bibfield  {author} {\bibinfo {author} {\bibfnamefont {C.~D.}\ \bibnamefont
  {Hill}}, \bibinfo {author} {\bibfnamefont {E.}~\bibnamefont {Peretz}},
  \bibinfo {author} {\bibfnamefont {S.~J.}\ \bibnamefont {Hile}}, \bibinfo
  {author} {\bibfnamefont {M.~G.}\ \bibnamefont {House}}, \bibinfo {author}
  {\bibfnamefont {M.}~\bibnamefont {Fuechsle}}, \bibinfo {author}
  {\bibfnamefont {S.}~\bibnamefont {Rogge}}, \bibinfo {author} {\bibfnamefont
  {M.~Y.}\ \bibnamefont {Simmons}}, \ and\ \bibinfo {author} {\bibfnamefont
  {L.~C.~L.}\ \bibnamefont {Hollenberg}},\ }\href {\doibase
  10.1126/sciadv.1500707} {\bibfield  {journal} {\bibinfo  {journal} {Science
  Advances}\ }\textbf {\bibinfo {volume} {1}},\ \bibinfo {pages} {e1500707}
  (\bibinfo {year} {2015})}\BibitemShut {NoStop}%
\bibitem [{\citenamefont {Veldhorst}\ \emph {et~al.}(2017)\citenamefont
  {Veldhorst}, \citenamefont {Eenink}, \citenamefont {Yang},\ and\
  \citenamefont {Dzurak}}]{veldhorst_silicon_2017}%
  \BibitemOpen
  \bibfield  {author} {\bibinfo {author} {\bibfnamefont {M.}~\bibnamefont
  {Veldhorst}}, \bibinfo {author} {\bibfnamefont {H.~G.~J.}\ \bibnamefont
  {Eenink}}, \bibinfo {author} {\bibfnamefont {C.~H.}\ \bibnamefont {Yang}}, \
  and\ \bibinfo {author} {\bibfnamefont {A.~S.}\ \bibnamefont {Dzurak}},\
  }\href {\doibase 10.1038/s41467-017-01905-6} {\bibfield  {journal} {\bibinfo
  {journal} {Nat. Commun.}\ }\textbf {\bibinfo {volume} {8}},\ \bibinfo {pages}
  {1766} (\bibinfo {year} {2017})}\BibitemShut {NoStop}%
\bibitem [{\citenamefont {Li}\ \emph {et~al.}(2017)\citenamefont {Li},
  \citenamefont {Petit}, \citenamefont {Franke}, \citenamefont {Dehollain},
  \citenamefont {Helsen}, \citenamefont {Steudtner}, \citenamefont {Thomas},
  \citenamefont {Yoscovits}, \citenamefont {Singh}, \citenamefont {Wehner},
  \citenamefont {Vandersypen}, \citenamefont {Clarke},\ and\ \citenamefont
  {Veldhorst}}]{li_crossbar_2017}%
  \BibitemOpen
  \bibfield  {author} {\bibinfo {author} {\bibfnamefont {R.}~\bibnamefont
  {Li}}, \bibinfo {author} {\bibfnamefont {L.}~\bibnamefont {Petit}}, \bibinfo
  {author} {\bibfnamefont {D.~P.}\ \bibnamefont {Franke}}, \bibinfo {author}
  {\bibfnamefont {J.~P.}\ \bibnamefont {Dehollain}}, \bibinfo {author}
  {\bibfnamefont {J.}~\bibnamefont {Helsen}}, \bibinfo {author} {\bibfnamefont
  {M.}~\bibnamefont {Steudtner}}, \bibinfo {author} {\bibfnamefont {N.~K.}\
  \bibnamefont {Thomas}}, \bibinfo {author} {\bibfnamefont {Z.~R.}\
  \bibnamefont {Yoscovits}}, \bibinfo {author} {\bibfnamefont {K.~J.}\
  \bibnamefont {Singh}}, \bibinfo {author} {\bibfnamefont {S.}~\bibnamefont
  {Wehner}}, \bibinfo {author} {\bibfnamefont {L.~M.~K.}\ \bibnamefont
  {Vandersypen}}, \bibinfo {author} {\bibfnamefont {J.~S.}\ \bibnamefont
  {Clarke}}, \ and\ \bibinfo {author} {\bibfnamefont {M.}~\bibnamefont
  {Veldhorst}},\ }\href {http://arxiv.org/abs/1711.03807} {\bibfield  {journal}
  {\bibinfo  {journal} {arXiv:1711.03807}\ } (\bibinfo {year}
  {2017})}\BibitemShut {NoStop}%
\bibitem [{\citenamefont {Hornibrook}\ \emph {et~al.}(2014)\citenamefont
  {Hornibrook}, \citenamefont {Colless}, \citenamefont {Mahoney}, \citenamefont
  {Croot}, \citenamefont {Blanvillain}, \citenamefont {Lu}, \citenamefont
  {Gossard},\ and\ \citenamefont {Reilly}}]{hornibrook_frequency_2014}%
  \BibitemOpen
  \bibfield  {author} {\bibinfo {author} {\bibfnamefont {J.~M.}\ \bibnamefont
  {Hornibrook}}, \bibinfo {author} {\bibfnamefont {J.~I.}\ \bibnamefont
  {Colless}}, \bibinfo {author} {\bibfnamefont {A.~C.}\ \bibnamefont
  {Mahoney}}, \bibinfo {author} {\bibfnamefont {X.~G.}\ \bibnamefont {Croot}},
  \bibinfo {author} {\bibfnamefont {S.}~\bibnamefont {Blanvillain}}, \bibinfo
  {author} {\bibfnamefont {H.}~\bibnamefont {Lu}}, \bibinfo {author}
  {\bibfnamefont {A.~C.}\ \bibnamefont {Gossard}}, \ and\ \bibinfo {author}
  {\bibfnamefont {D.~J.}\ \bibnamefont {Reilly}},\ }\href {\doibase
  10.1063/1.4868107} {\bibfield  {journal} {\bibinfo  {journal} {Appl. Phys.
  Lett.}\ }\textbf {\bibinfo {volume} {104}},\ \bibinfo {pages} {103108}
  (\bibinfo {year} {2014})}\BibitemShut {NoStop}%
\bibitem [{\citenamefont {Homulle}\ \emph {et~al.}(2016)\citenamefont
  {Homulle}, \citenamefont {Visser}, \citenamefont {Patra}, \citenamefont
  {Ferrari}, \citenamefont {Prati}, \citenamefont {Almudever}, \citenamefont
  {Bertels}, \citenamefont {Sebastiano},\ and\ \citenamefont
  {Charbon}}]{homulle_cryocmos_2016}%
  \BibitemOpen
  \bibfield  {author} {\bibinfo {author} {\bibfnamefont {H.}~\bibnamefont
  {Homulle}}, \bibinfo {author} {\bibfnamefont {S.}~\bibnamefont {Visser}},
  \bibinfo {author} {\bibfnamefont {B.}~\bibnamefont {Patra}}, \bibinfo
  {author} {\bibfnamefont {G.}~\bibnamefont {Ferrari}}, \bibinfo {author}
  {\bibfnamefont {E.}~\bibnamefont {Prati}}, \bibinfo {author} {\bibfnamefont
  {C.~G.}\ \bibnamefont {Almudever}}, \bibinfo {author} {\bibfnamefont
  {K.}~\bibnamefont {Bertels}}, \bibinfo {author} {\bibfnamefont
  {F.}~\bibnamefont {Sebastiano}}, \ and\ \bibinfo {author} {\bibfnamefont
  {E.}~\bibnamefont {Charbon}},\ }in\ \href {\doibase 10.1145/2903150.2906828}
  {\emph {\bibinfo {booktitle} {Proceedings of the {ACM} {International}
  {Conference} on {Computing} {Frontiers}}}}\ (\bibinfo  {publisher} {ACM},\
  \bibinfo {address} {New York},\ \bibinfo {year} {2016})\ pp.\ \bibinfo
  {pages} {282--287}\BibitemShut {NoStop}%
\bibitem [{\citenamefont {Vandersypen}\ \emph {et~al.}(2017)\citenamefont
  {Vandersypen}, \citenamefont {Bluhm}, \citenamefont {Clarke}, \citenamefont
  {Dzurak}, \citenamefont {Ishihara}, \citenamefont {Morello}, \citenamefont
  {Reilly}, \citenamefont {Schreiber},\ and\ \citenamefont
  {Veldhorst}}]{vandersypen_interfacing_2017}%
  \BibitemOpen
  \bibfield  {author} {\bibinfo {author} {\bibfnamefont {L.~M.~K.}\
  \bibnamefont {Vandersypen}}, \bibinfo {author} {\bibfnamefont
  {H.}~\bibnamefont {Bluhm}}, \bibinfo {author} {\bibfnamefont {J.~S.}\
  \bibnamefont {Clarke}}, \bibinfo {author} {\bibfnamefont {A.~S.}\
  \bibnamefont {Dzurak}}, \bibinfo {author} {\bibfnamefont {R.}~\bibnamefont
  {Ishihara}}, \bibinfo {author} {\bibfnamefont {A.}~\bibnamefont {Morello}},
  \bibinfo {author} {\bibfnamefont {D.~J.}\ \bibnamefont {Reilly}}, \bibinfo
  {author} {\bibfnamefont {L.~R.}\ \bibnamefont {Schreiber}}, \ and\ \bibinfo
  {author} {\bibfnamefont {M.}~\bibnamefont {Veldhorst}},\ }\href {\doibase
  10.1038/s41534-017-0038-y} {\bibfield  {journal} {\bibinfo  {journal} {npj
  Quantum Information}\ }\textbf {\bibinfo {volume} {3}},\ \bibinfo {pages}
  {34} (\bibinfo {year} {2017})}\BibitemShut {NoStop}%
\bibitem [{\citenamefont {Petit}\ \emph {et~al.}(2018)\citenamefont {Petit},
  \citenamefont {Boter}, \citenamefont {Eenink}, \citenamefont {Droulers},
  \citenamefont {Tagliaferri}, \citenamefont {Li}, \citenamefont {Franke},
  \citenamefont {Singh}, \citenamefont {Clarke}, \citenamefont {Schouten},
  \citenamefont {Dobrovitski}, \citenamefont {Vandersypen},\ and\ \citenamefont
  {Veldhorst}}]{petit_spin_2018}%
  \BibitemOpen
  \bibfield  {author} {\bibinfo {author} {\bibfnamefont {L.}~\bibnamefont
  {Petit}}, \bibinfo {author} {\bibfnamefont {J.~M.}\ \bibnamefont {Boter}},
  \bibinfo {author} {\bibfnamefont {H.~G.~J.}\ \bibnamefont {Eenink}}, \bibinfo
  {author} {\bibfnamefont {G.}~\bibnamefont {Droulers}}, \bibinfo {author}
  {\bibfnamefont {M.~L.~V.}\ \bibnamefont {Tagliaferri}}, \bibinfo {author}
  {\bibfnamefont {R.}~\bibnamefont {Li}}, \bibinfo {author} {\bibfnamefont
  {D.~P.}\ \bibnamefont {Franke}}, \bibinfo {author} {\bibfnamefont {K.~J.}\
  \bibnamefont {Singh}}, \bibinfo {author} {\bibfnamefont {J.~S.}\ \bibnamefont
  {Clarke}}, \bibinfo {author} {\bibfnamefont {R.~N.}\ \bibnamefont
  {Schouten}}, \bibinfo {author} {\bibfnamefont {V.~V.}\ \bibnamefont
  {Dobrovitski}}, \bibinfo {author} {\bibfnamefont {L.~M.~K.}\ \bibnamefont
  {Vandersypen}}, \ and\ \bibinfo {author} {\bibfnamefont {M.}~\bibnamefont
  {Veldhorst}},\ }\href {http://arxiv.org/abs/1803.01774} {\bibfield  {journal}
  {\bibinfo  {journal} {arXiv:1803.01774}\ } (\bibinfo {year}
  {2018})}\BibitemShut {NoStop}%
\bibitem [{\citenamefont {Helsen}\ \emph {et~al.}(2018)\citenamefont {Helsen},
  \citenamefont {Steudtner}, \citenamefont {Veldhorst},\ and\ \citenamefont
  {Wehner}}]{helsen_quantum_2018}%
  \BibitemOpen
  \bibfield  {author} {\bibinfo {author} {\bibfnamefont {J.}~\bibnamefont
  {Helsen}}, \bibinfo {author} {\bibfnamefont {M.}~\bibnamefont {Steudtner}},
  \bibinfo {author} {\bibfnamefont {M.}~\bibnamefont {Veldhorst}}, \ and\
  \bibinfo {author} {\bibfnamefont {S.}~\bibnamefont {Wehner}},\ }\href
  {\doibase 10.1088/2058-9565/aab8b0} {\bibfield  {journal} {\bibinfo
  {journal} {Quantum Sci. Technol.}\ }\textbf {\bibinfo {volume} {3}},\
  \bibinfo {pages} {035005} (\bibinfo {year} {2018})}\BibitemShut {NoStop}%
\bibitem [{\citenamefont {Bishop}\ \emph {et~al.}(2017)\citenamefont {Bishop},
  \citenamefont {Bravyi}, \citenamefont {Cross}, \citenamefont {Gambetta},\
  and\ \citenamefont {Smolin}}]{bishop_quantum_2017}%
  \BibitemOpen
  \bibfield  {author} {\bibinfo {author} {\bibfnamefont {L.~S.}\ \bibnamefont
  {Bishop}}, \bibinfo {author} {\bibfnamefont {S.}~\bibnamefont {Bravyi}},
  \bibinfo {author} {\bibfnamefont {A.}~\bibnamefont {Cross}}, \bibinfo
  {author} {\bibfnamefont {J.~M.}\ \bibnamefont {Gambetta}}, \ and\ \bibinfo
  {author} {\bibfnamefont {J.}~\bibnamefont {Smolin}},\ }\href
  {https://dal.objectstorage.open.softlayer.com/v1/AUTH_039c3bf6e6e54d76b8e66152e2f87877/community-documents/quatnum-volumehp08co1vbo0cc8fr.pdf}
  {\enquote {\bibinfo {title} {Quantum {Volume}},}\ } (\bibinfo {year}
  {2017})\BibitemShut {NoStop}%
\bibitem [{\citenamefont {Moll}\ \emph {et~al.}(2017)\citenamefont {Moll},
  \citenamefont {Barkoutsos}, \citenamefont {Bishop}, \citenamefont {Chow},
  \citenamefont {Cross}, \citenamefont {Egger}, \citenamefont {Filipp},
  \citenamefont {Fuhrer}, \citenamefont {Gambetta}, \citenamefont {Ganzhorn},
  \citenamefont {Kandala}, \citenamefont {Mezzacapo}, \citenamefont {Müller},
  \citenamefont {Riess}, \citenamefont {Salis}, \citenamefont {Smolin},
  \citenamefont {Tavernelli},\ and\ \citenamefont {Temme}}]{moll_quantum_2017}%
  \BibitemOpen
  \bibfield  {author} {\bibinfo {author} {\bibfnamefont {N.}~\bibnamefont
  {Moll}}, \bibinfo {author} {\bibfnamefont {P.}~\bibnamefont {Barkoutsos}},
  \bibinfo {author} {\bibfnamefont {L.~S.}\ \bibnamefont {Bishop}}, \bibinfo
  {author} {\bibfnamefont {J.~M.}\ \bibnamefont {Chow}}, \bibinfo {author}
  {\bibfnamefont {A.}~\bibnamefont {Cross}}, \bibinfo {author} {\bibfnamefont
  {D.~J.}\ \bibnamefont {Egger}}, \bibinfo {author} {\bibfnamefont
  {S.}~\bibnamefont {Filipp}}, \bibinfo {author} {\bibfnamefont
  {A.}~\bibnamefont {Fuhrer}}, \bibinfo {author} {\bibfnamefont {J.~M.}\
  \bibnamefont {Gambetta}}, \bibinfo {author} {\bibfnamefont {M.}~\bibnamefont
  {Ganzhorn}}, \bibinfo {author} {\bibfnamefont {A.}~\bibnamefont {Kandala}},
  \bibinfo {author} {\bibfnamefont {A.}~\bibnamefont {Mezzacapo}}, \bibinfo
  {author} {\bibfnamefont {P.}~\bibnamefont {Müller}}, \bibinfo {author}
  {\bibfnamefont {W.}~\bibnamefont {Riess}}, \bibinfo {author} {\bibfnamefont
  {G.}~\bibnamefont {Salis}}, \bibinfo {author} {\bibfnamefont
  {J.}~\bibnamefont {Smolin}}, \bibinfo {author} {\bibfnamefont
  {I.}~\bibnamefont {Tavernelli}}, \ and\ \bibinfo {author} {\bibfnamefont
  {K.}~\bibnamefont {Temme}},\ }\href {http://arxiv.org/abs/1710.01022}
  {\bibfield  {journal} {\bibinfo  {journal} {arXiv:1710.01022}\ } (\bibinfo
  {year} {2017})}\BibitemShut {NoStop}%
\bibitem [{\citenamefont {Preskill}(2012)}]{preskill_quantum_2012}%
  \BibitemOpen
  \bibfield  {author} {\bibinfo {author} {\bibfnamefont {J.}~\bibnamefont
  {Preskill}},\ }\href {http://arxiv.org/abs/1203.5813} {\bibfield  {journal}
  {\bibinfo  {journal} {arXiv:1203.5813}\ } (\bibinfo {year}
  {2012})}\BibitemShut {NoStop}%
\bibitem [{\citenamefont {Boixo}\ \emph {et~al.}(2018)\citenamefont {Boixo},
  \citenamefont {Isakov}, \citenamefont {Smelyanskiy}, \citenamefont {Babbush},
  \citenamefont {Ding}, \citenamefont {Jiang}, \citenamefont {Bremner},
  \citenamefont {Martinis},\ and\ \citenamefont
  {Neven}}]{boixo_characterizing_2018}%
  \BibitemOpen
  \bibfield  {author} {\bibinfo {author} {\bibfnamefont {S.}~\bibnamefont
  {Boixo}}, \bibinfo {author} {\bibfnamefont {S.~V.}\ \bibnamefont {Isakov}},
  \bibinfo {author} {\bibfnamefont {V.~N.}\ \bibnamefont {Smelyanskiy}},
  \bibinfo {author} {\bibfnamefont {R.}~\bibnamefont {Babbush}}, \bibinfo
  {author} {\bibfnamefont {N.}~\bibnamefont {Ding}}, \bibinfo {author}
  {\bibfnamefont {Z.}~\bibnamefont {Jiang}}, \bibinfo {author} {\bibfnamefont
  {M.~J.}\ \bibnamefont {Bremner}}, \bibinfo {author} {\bibfnamefont {J.~M.}\
  \bibnamefont {Martinis}}, \ and\ \bibinfo {author} {\bibfnamefont
  {H.}~\bibnamefont {Neven}},\ }\href {\doibase 10.1038/s41567-018-0124-x}
  {\bibfield  {journal} {\bibinfo  {journal} {Nat. Phys.}\ }\textbf {\bibinfo
  {volume} {14}},\ \bibinfo {pages} {595} (\bibinfo {year} {2018})}\BibitemShut
  {NoStop}%
\bibitem [{\citenamefont {Pednault}\ \emph {et~al.}(2017)\citenamefont
  {Pednault}, \citenamefont {Gunnels}, \citenamefont {Nannicini}, \citenamefont
  {Horesh}, \citenamefont {Magerlein}, \citenamefont {Solomonik},\ and\
  \citenamefont {Wisnieff}}]{pednault_breaking_2017}%
  \BibitemOpen
  \bibfield  {author} {\bibinfo {author} {\bibfnamefont {E.}~\bibnamefont
  {Pednault}}, \bibinfo {author} {\bibfnamefont {J.~A.}\ \bibnamefont
  {Gunnels}}, \bibinfo {author} {\bibfnamefont {G.}~\bibnamefont {Nannicini}},
  \bibinfo {author} {\bibfnamefont {L.}~\bibnamefont {Horesh}}, \bibinfo
  {author} {\bibfnamefont {T.}~\bibnamefont {Magerlein}}, \bibinfo {author}
  {\bibfnamefont {E.}~\bibnamefont {Solomonik}}, \ and\ \bibinfo {author}
  {\bibfnamefont {R.}~\bibnamefont {Wisnieff}},\ }\href
  {http://arxiv.org/abs/1710.05867} {\bibfield  {journal} {\bibinfo  {journal}
  {arXiv:1710.05867}\ } (\bibinfo {year} {2017})}\BibitemShut {NoStop}%
\bibitem [{\citenamefont {Chen}\ \emph {et~al.}(2018)\citenamefont {Chen},
  \citenamefont {Zhang}, \citenamefont {Huang}, \citenamefont {Newman},\ and\
  \citenamefont {Shi}}]{chen_classical_2018}%
  \BibitemOpen
  \bibfield  {author} {\bibinfo {author} {\bibfnamefont {J.}~\bibnamefont
  {Chen}}, \bibinfo {author} {\bibfnamefont {F.}~\bibnamefont {Zhang}},
  \bibinfo {author} {\bibfnamefont {C.}~\bibnamefont {Huang}}, \bibinfo
  {author} {\bibfnamefont {M.}~\bibnamefont {Newman}}, \ and\ \bibinfo {author}
  {\bibfnamefont {Y.}~\bibnamefont {Shi}},\ }\href
  {http://arxiv.org/abs/1805.01450} {\bibfield  {journal} {\bibinfo  {journal}
  {arXiv:1805.01450}\ } (\bibinfo {year} {2018})}\BibitemShut {NoStop}%
\bibitem [{\citenamefont {Harrow}\ and\ \citenamefont
  {Montanaro}(2017)}]{harrow_quantum_2017}%
  \BibitemOpen
  \bibfield  {author} {\bibinfo {author} {\bibfnamefont {A.~W.}\ \bibnamefont
  {Harrow}}\ and\ \bibinfo {author} {\bibfnamefont {A.}~\bibnamefont
  {Montanaro}},\ }\href {\doibase 10.1038/nature23458} {\bibfield  {journal}
  {\bibinfo  {journal} {Nature}\ }\textbf {\bibinfo {volume} {549}},\ \bibinfo
  {pages} {203} (\bibinfo {year} {2017})}\BibitemShut {NoStop}%
\bibitem [{\citenamefont {Preskill}(2018)}]{preskill_quantum_2018}%
  \BibitemOpen
  \bibfield  {author} {\bibinfo {author} {\bibfnamefont {J.}~\bibnamefont
  {Preskill}},\ }\href {http://arxiv.org/abs/1801.00862} {\bibfield  {journal}
  {\bibinfo  {journal} {arXiv:1801.00862}\ } (\bibinfo {year}
  {2018})}\BibitemShut {NoStop}%
\bibitem [{\citenamefont {Lloyd}(1996)}]{lloyd_universal_1996}%
  \BibitemOpen
  \bibfield  {author} {\bibinfo {author} {\bibfnamefont {S.}~\bibnamefont
  {Lloyd}},\ }\href {\doibase 10.1126/science.273.5278.1073} {\bibfield
  {journal} {\bibinfo  {journal} {Science}\ }\textbf {\bibinfo {volume}
  {273}},\ \bibinfo {pages} {1073} (\bibinfo {year} {1996})}\BibitemShut
  {NoStop}%
\bibitem [{\citenamefont {Cirac}\ and\ \citenamefont
  {Zoller}(2012)}]{cirac_goals_2012}%
  \BibitemOpen
  \bibfield  {author} {\bibinfo {author} {\bibfnamefont {J.~I.}\ \bibnamefont
  {Cirac}}\ and\ \bibinfo {author} {\bibfnamefont {P.}~\bibnamefont {Zoller}},\
  }\href {\doibase 10.1038/nphys2275} {\bibfield  {journal} {\bibinfo
  {journal} {Nat. Phys.}\ }\textbf {\bibinfo {volume} {8}},\ \bibinfo {pages}
  {264} (\bibinfo {year} {2012})}\BibitemShut {NoStop}%
\bibitem [{\citenamefont {Jones}\ \emph {et~al.}(2012)\citenamefont {Jones},
  \citenamefont {Van~Meter}, \citenamefont {Fowler}, \citenamefont {McMahon},
  \citenamefont {Kim}, \citenamefont {Ladd},\ and\ \citenamefont
  {Yamamoto}}]{jones_layered_2012}%
  \BibitemOpen
  \bibfield  {author} {\bibinfo {author} {\bibfnamefont {N.~C.}\ \bibnamefont
  {Jones}}, \bibinfo {author} {\bibfnamefont {R.}~\bibnamefont {Van~Meter}},
  \bibinfo {author} {\bibfnamefont {A.~G.}\ \bibnamefont {Fowler}}, \bibinfo
  {author} {\bibfnamefont {P.~L.}\ \bibnamefont {McMahon}}, \bibinfo {author}
  {\bibfnamefont {J.}~\bibnamefont {Kim}}, \bibinfo {author} {\bibfnamefont
  {T.~D.}\ \bibnamefont {Ladd}}, \ and\ \bibinfo {author} {\bibfnamefont
  {Y.}~\bibnamefont {Yamamoto}},\ }\href {\doibase 10.1103/PhysRevX.2.031007}
  {\bibfield  {journal} {\bibinfo  {journal} {Phys. Rev. X}\ }\textbf {\bibinfo
  {volume} {2}},\ \bibinfo {pages} {031007} (\bibinfo {year}
  {2012})}\BibitemShut {NoStop}%
\bibitem [{\citenamefont {Fu}\ \emph {et~al.}(2016)\citenamefont {Fu},
  \citenamefont {Riesebos}, \citenamefont {Lao}, \citenamefont {Almudever},
  \citenamefont {Sebastiano}, \citenamefont {Versluis}, \citenamefont
  {Charbon},\ and\ \citenamefont {Bertels}}]{fu_heterogeneous_2016}%
  \BibitemOpen
  \bibfield  {author} {\bibinfo {author} {\bibfnamefont {X.}~\bibnamefont
  {Fu}}, \bibinfo {author} {\bibfnamefont {L.}~\bibnamefont {Riesebos}},
  \bibinfo {author} {\bibfnamefont {L.}~\bibnamefont {Lao}}, \bibinfo {author}
  {\bibfnamefont {C.~G.}\ \bibnamefont {Almudever}}, \bibinfo {author}
  {\bibfnamefont {F.}~\bibnamefont {Sebastiano}}, \bibinfo {author}
  {\bibfnamefont {R.}~\bibnamefont {Versluis}}, \bibinfo {author}
  {\bibfnamefont {E.}~\bibnamefont {Charbon}}, \ and\ \bibinfo {author}
  {\bibfnamefont {K.}~\bibnamefont {Bertels}},\ }in\ \href {\doibase
  10.1145/2903150.2906827} {\emph {\bibinfo {booktitle} {Proceedings of the
  {ACM} {International} {Conference} on {Computing} {Frontiers}}}}\ (\bibinfo
  {publisher} {ACM},\ \bibinfo {address} {New York},\ \bibinfo {year} {2016})\
  pp.\ \bibinfo {pages} {323--330}\BibitemShut {NoStop}%
\bibitem [{\citenamefont {Feynman}(1982)}]{feynman_simulating_1982}%
  \BibitemOpen
  \bibfield  {author} {\bibinfo {author} {\bibfnamefont {R.~P.}\ \bibnamefont
  {Feynman}},\ }\href {\doibase 10.1007/BF02650179} {\bibfield  {journal}
  {\bibinfo  {journal} {Int. J. Theor. Phys.}\ }\textbf {\bibinfo {volume}
  {21}},\ \bibinfo {pages} {467} (\bibinfo {year} {1982})}\BibitemShut
  {NoStop}%
\bibitem [{\citenamefont {Ladd}\ \emph {et~al.}(2010)\citenamefont {Ladd},
  \citenamefont {Jelezko}, \citenamefont {Laflamme}, \citenamefont {Nakamura},
  \citenamefont {Monroe},\ and\ \citenamefont {O’Brien}}]{ladd_quantum_2010}%
  \BibitemOpen
  \bibfield  {author} {\bibinfo {author} {\bibfnamefont {T.~D.}\ \bibnamefont
  {Ladd}}, \bibinfo {author} {\bibfnamefont {F.}~\bibnamefont {Jelezko}},
  \bibinfo {author} {\bibfnamefont {R.}~\bibnamefont {Laflamme}}, \bibinfo
  {author} {\bibfnamefont {Y.}~\bibnamefont {Nakamura}}, \bibinfo {author}
  {\bibfnamefont {C.}~\bibnamefont {Monroe}}, \ and\ \bibinfo {author}
  {\bibfnamefont {J.~L.}\ \bibnamefont {O’Brien}},\ }\href {\doibase
  10.1038/nature08812} {\bibfield  {journal} {\bibinfo  {journal} {Nature}\
  }\textbf {\bibinfo {volume} {464}},\ \bibinfo {pages} {45} (\bibinfo {year}
  {2010})}\BibitemShut {NoStop}%
\bibitem [{\citenamefont {Hollenberg}\ \emph {et~al.}(2006)\citenamefont
  {Hollenberg}, \citenamefont {Greentree}, \citenamefont {Fowler},\ and\
  \citenamefont {Wellard}}]{hollenberg_two-dimensional_2006}%
  \BibitemOpen
  \bibfield  {author} {\bibinfo {author} {\bibfnamefont {L.~C.~L.}\
  \bibnamefont {Hollenberg}}, \bibinfo {author} {\bibfnamefont {A.~D.}\
  \bibnamefont {Greentree}}, \bibinfo {author} {\bibfnamefont {A.~G.}\
  \bibnamefont {Fowler}}, \ and\ \bibinfo {author} {\bibfnamefont {C.~J.}\
  \bibnamefont {Wellard}},\ }\href {\doibase 10.1103/PhysRevB.74.045311}
  {\bibfield  {journal} {\bibinfo  {journal} {Phys. Rev. B}\ }\textbf {\bibinfo
  {volume} {74}},\ \bibinfo {pages} {045311} (\bibinfo {year}
  {2006})}\BibitemShut {NoStop}%
\bibitem [{\citenamefont {Watson}\ \emph {et~al.}(2018)\citenamefont {Watson},
  \citenamefont {Philips}, \citenamefont {Kawakami}, \citenamefont {Ward},
  \citenamefont {Scarlino}, \citenamefont {Veldhorst}, \citenamefont {Savage},
  \citenamefont {Lagally}, \citenamefont {Friesen}, \citenamefont
  {Coppersmith}, \citenamefont {Eriksson},\ and\ \citenamefont
  {Vandersypen}}]{watson_programmable_2018}%
  \BibitemOpen
  \bibfield  {author} {\bibinfo {author} {\bibfnamefont {T.~F.}\ \bibnamefont
  {Watson}}, \bibinfo {author} {\bibfnamefont {S.~G.~J.}\ \bibnamefont
  {Philips}}, \bibinfo {author} {\bibfnamefont {E.}~\bibnamefont {Kawakami}},
  \bibinfo {author} {\bibfnamefont {D.~R.}\ \bibnamefont {Ward}}, \bibinfo
  {author} {\bibfnamefont {P.}~\bibnamefont {Scarlino}}, \bibinfo {author}
  {\bibfnamefont {M.}~\bibnamefont {Veldhorst}}, \bibinfo {author}
  {\bibfnamefont {D.~E.}\ \bibnamefont {Savage}}, \bibinfo {author}
  {\bibfnamefont {M.~G.}\ \bibnamefont {Lagally}}, \bibinfo {author}
  {\bibfnamefont {M.}~\bibnamefont {Friesen}}, \bibinfo {author} {\bibfnamefont
  {S.~N.}\ \bibnamefont {Coppersmith}}, \bibinfo {author} {\bibfnamefont
  {M.~A.}\ \bibnamefont {Eriksson}}, \ and\ \bibinfo {author} {\bibfnamefont
  {L.~M.~K.}\ \bibnamefont {Vandersypen}},\ }\href {\doibase
  10.1038/nature25766} {\bibfield  {journal} {\bibinfo  {journal} {Nature}\
  }\textbf {\bibinfo {volume} {555}},\ \bibinfo {pages} {633} (\bibinfo {year}
  {2018})}\BibitemShut {NoStop}%
\bibitem [{\citenamefont {Devoret}\ and\ \citenamefont
  {Schoelkopf}(2013)}]{devoret_superconducting_2013}%
  \BibitemOpen
  \bibfield  {author} {\bibinfo {author} {\bibfnamefont {M.~H.}\ \bibnamefont
  {Devoret}}\ and\ \bibinfo {author} {\bibfnamefont {R.~J.}\ \bibnamefont
  {Schoelkopf}},\ }\href {\doibase 10.1126/science.1231930} {\bibfield
  {journal} {\bibinfo  {journal} {Science}\ }\textbf {\bibinfo {volume}
  {339}},\ \bibinfo {pages} {1169} (\bibinfo {year} {2013})}\BibitemShut
  {NoStop}%
\bibitem [{noa(2017)}]{noauthor_ibm_2017}%
  \BibitemOpen
  \href {https://www-03.ibm.com/press/us/en/pressrelease/53374.wss} {\enquote
  {\bibinfo {title} {{IBM} {Announces} {Advances} to {IBM} {Q} {Systems} \&
  {Ecosystem}},}\ } (\bibinfo {year} {2017}),\ \bibinfo {note}
  {https://www-03.ibm.com/press/us/en/pressrelease/53374.wss}\BibitemShut
  {NoStop}%
\bibitem [{noa(2018)}]{noauthor_2018_2018}%
  \BibitemOpen
  \href
  {https://newsroom.intel.com/news/intel-advances-quantum-neuromorphic-computing-research/}
  {\enquote {\bibinfo {title} {2018 {CES}: {Intel} {Advances} {Quantum} and
  {Neuromorphic} {Computing} {Research}},}\ } (\bibinfo {year} {2018}),\
  \bibinfo {note}
  {https://newsroom.intel.com/news/intel-advances-quantum-neuromorphic-computing-research/}\BibitemShut
  {NoStop}%
\bibitem [{\citenamefont {Neill}\ \emph {et~al.}(2018)\citenamefont {Neill},
  \citenamefont {Roushan}, \citenamefont {Kechedzhi}, \citenamefont {Boixo},
  \citenamefont {Isakov}, \citenamefont {Smelyanskiy}, \citenamefont {Megrant},
  \citenamefont {Chiaro}, \citenamefont {Dunsworth}, \citenamefont {Arya},
  \citenamefont {Barends}, \citenamefont {Burkett}, \citenamefont {Chen},
  \citenamefont {Chen}, \citenamefont {Fowler}, \citenamefont {Foxen},
  \citenamefont {Giustina}, \citenamefont {Graff}, \citenamefont {Jeffrey},
  \citenamefont {Huang}, \citenamefont {Kelly}, \citenamefont {Klimov},
  \citenamefont {Lucero}, \citenamefont {Mutus}, \citenamefont {Neeley},
  \citenamefont {Quintana}, \citenamefont {Sank}, \citenamefont {Vainsencher},
  \citenamefont {Wenner}, \citenamefont {White}, \citenamefont {Neven},\ and\
  \citenamefont {Martinis}}]{neill_blueprint_2018}%
  \BibitemOpen
  \bibfield  {author} {\bibinfo {author} {\bibfnamefont {C.}~\bibnamefont
  {Neill}}, \bibinfo {author} {\bibfnamefont {P.}~\bibnamefont {Roushan}},
  \bibinfo {author} {\bibfnamefont {K.}~\bibnamefont {Kechedzhi}}, \bibinfo
  {author} {\bibfnamefont {S.}~\bibnamefont {Boixo}}, \bibinfo {author}
  {\bibfnamefont {S.~V.}\ \bibnamefont {Isakov}}, \bibinfo {author}
  {\bibfnamefont {V.}~\bibnamefont {Smelyanskiy}}, \bibinfo {author}
  {\bibfnamefont {A.}~\bibnamefont {Megrant}}, \bibinfo {author} {\bibfnamefont
  {B.}~\bibnamefont {Chiaro}}, \bibinfo {author} {\bibfnamefont
  {A.}~\bibnamefont {Dunsworth}}, \bibinfo {author} {\bibfnamefont
  {K.}~\bibnamefont {Arya}}, \bibinfo {author} {\bibfnamefont {R.}~\bibnamefont
  {Barends}}, \bibinfo {author} {\bibfnamefont {B.}~\bibnamefont {Burkett}},
  \bibinfo {author} {\bibfnamefont {Y.}~\bibnamefont {Chen}}, \bibinfo {author}
  {\bibfnamefont {Z.}~\bibnamefont {Chen}}, \bibinfo {author} {\bibfnamefont
  {A.}~\bibnamefont {Fowler}}, \bibinfo {author} {\bibfnamefont
  {B.}~\bibnamefont {Foxen}}, \bibinfo {author} {\bibfnamefont
  {M.}~\bibnamefont {Giustina}}, \bibinfo {author} {\bibfnamefont
  {R.}~\bibnamefont {Graff}}, \bibinfo {author} {\bibfnamefont
  {E.}~\bibnamefont {Jeffrey}}, \bibinfo {author} {\bibfnamefont
  {T.}~\bibnamefont {Huang}}, \bibinfo {author} {\bibfnamefont
  {J.}~\bibnamefont {Kelly}}, \bibinfo {author} {\bibfnamefont
  {P.}~\bibnamefont {Klimov}}, \bibinfo {author} {\bibfnamefont
  {E.}~\bibnamefont {Lucero}}, \bibinfo {author} {\bibfnamefont
  {J.}~\bibnamefont {Mutus}}, \bibinfo {author} {\bibfnamefont
  {M.}~\bibnamefont {Neeley}}, \bibinfo {author} {\bibfnamefont
  {C.}~\bibnamefont {Quintana}}, \bibinfo {author} {\bibfnamefont
  {D.}~\bibnamefont {Sank}}, \bibinfo {author} {\bibfnamefont {A.}~\bibnamefont
  {Vainsencher}}, \bibinfo {author} {\bibfnamefont {J.}~\bibnamefont {Wenner}},
  \bibinfo {author} {\bibfnamefont {T.~C.}\ \bibnamefont {White}}, \bibinfo
  {author} {\bibfnamefont {H.}~\bibnamefont {Neven}}, \ and\ \bibinfo {author}
  {\bibfnamefont {J.~M.}\ \bibnamefont {Martinis}},\ }\href {\doibase
  10.1126/science.aao4309} {\bibfield  {journal} {\bibinfo  {journal}
  {Science}\ }\textbf {\bibinfo {volume} {360}},\ \bibinfo {pages} {195}
  (\bibinfo {year} {2018})}\BibitemShut {NoStop}%
\bibitem [{\citenamefont {Kelly}(2018)}]{kelly_preview_2018}%
  \BibitemOpen
  \bibfield  {author} {\bibinfo {author} {\bibfnamefont {J.}~\bibnamefont
  {Kelly}},\ }\href
  {https://research.googleblog.com/2018/03/a-preview-of-bristlecone-googles-new.html}
  {\enquote {\bibinfo {title} {A {Preview} of {Bristlecone}, {Google}’s {New}
  {Quantum} {Processor}},}\ } (\bibinfo {year} {2018}),\ \bibinfo {note}
  {https://research.googleblog.com/2018/03/a-preview-of-bristlecone-googles-new.html}\BibitemShut
  {NoStop}%
\bibitem [{\citenamefont {Versluis}\ \emph {et~al.}(2017)\citenamefont
  {Versluis}, \citenamefont {Poletto}, \citenamefont {Khammassi}, \citenamefont
  {Tarasinski}, \citenamefont {Haider}, \citenamefont {Michalak}, \citenamefont
  {Bruno}, \citenamefont {Bertels},\ and\ \citenamefont
  {DiCarlo}}]{versluis_scalable_2017}%
  \BibitemOpen
  \bibfield  {author} {\bibinfo {author} {\bibfnamefont {R.}~\bibnamefont
  {Versluis}}, \bibinfo {author} {\bibfnamefont {S.}~\bibnamefont {Poletto}},
  \bibinfo {author} {\bibfnamefont {N.}~\bibnamefont {Khammassi}}, \bibinfo
  {author} {\bibfnamefont {B.}~\bibnamefont {Tarasinski}}, \bibinfo {author}
  {\bibfnamefont {N.}~\bibnamefont {Haider}}, \bibinfo {author} {\bibfnamefont
  {D.}~\bibnamefont {Michalak}}, \bibinfo {author} {\bibfnamefont
  {A.}~\bibnamefont {Bruno}}, \bibinfo {author} {\bibfnamefont
  {K.}~\bibnamefont {Bertels}}, \ and\ \bibinfo {author} {\bibfnamefont
  {L.}~\bibnamefont {DiCarlo}},\ }\href {\doibase
  10.1103/PhysRevApplied.8.034021} {\bibfield  {journal} {\bibinfo  {journal}
  {Phys. Rev. Applied}\ }\textbf {\bibinfo {volume} {8}},\ \bibinfo {pages}
  {034021} (\bibinfo {year} {2017})}\BibitemShut {NoStop}%
\bibitem [{\citenamefont {Monz}\ \emph {et~al.}(2011)\citenamefont {Monz},
  \citenamefont {Schindler}, \citenamefont {Barreiro}, \citenamefont {Chwalla},
  \citenamefont {Nigg}, \citenamefont {Coish}, \citenamefont {Harlander},
  \citenamefont {Hänsel}, \citenamefont {Hennrich},\ and\ \citenamefont
  {Blatt}}]{monz_14-qubit_2011}%
  \BibitemOpen
  \bibfield  {author} {\bibinfo {author} {\bibfnamefont {T.}~\bibnamefont
  {Monz}}, \bibinfo {author} {\bibfnamefont {P.}~\bibnamefont {Schindler}},
  \bibinfo {author} {\bibfnamefont {J.~T.}\ \bibnamefont {Barreiro}}, \bibinfo
  {author} {\bibfnamefont {M.}~\bibnamefont {Chwalla}}, \bibinfo {author}
  {\bibfnamefont {D.}~\bibnamefont {Nigg}}, \bibinfo {author} {\bibfnamefont
  {W.~A.}\ \bibnamefont {Coish}}, \bibinfo {author} {\bibfnamefont
  {M.}~\bibnamefont {Harlander}}, \bibinfo {author} {\bibfnamefont
  {W.}~\bibnamefont {Hänsel}}, \bibinfo {author} {\bibfnamefont
  {M.}~\bibnamefont {Hennrich}}, \ and\ \bibinfo {author} {\bibfnamefont
  {R.}~\bibnamefont {Blatt}},\ }\href {\doibase 10.1103/PhysRevLett.106.130506}
  {\bibfield  {journal} {\bibinfo  {journal} {Phys. Rev. Lett.}\ }\textbf
  {\bibinfo {volume} {106}},\ \bibinfo {pages} {130506} (\bibinfo {year}
  {2011})}\BibitemShut {NoStop}%
\bibitem [{\citenamefont {Friis}\ \emph {et~al.}(2018)\citenamefont {Friis},
  \citenamefont {Marty}, \citenamefont {Maier}, \citenamefont {Hempel},
  \citenamefont {Holzäpfel}, \citenamefont {Jurcevic}, \citenamefont {Plenio},
  \citenamefont {Huber}, \citenamefont {Roos}, \citenamefont {Blatt},\ and\
  \citenamefont {Lanyon}}]{friis_observation_2018}%
  \BibitemOpen
  \bibfield  {author} {\bibinfo {author} {\bibfnamefont {N.}~\bibnamefont
  {Friis}}, \bibinfo {author} {\bibfnamefont {O.}~\bibnamefont {Marty}},
  \bibinfo {author} {\bibfnamefont {C.}~\bibnamefont {Maier}}, \bibinfo
  {author} {\bibfnamefont {C.}~\bibnamefont {Hempel}}, \bibinfo {author}
  {\bibfnamefont {M.}~\bibnamefont {Holzäpfel}}, \bibinfo {author}
  {\bibfnamefont {P.}~\bibnamefont {Jurcevic}}, \bibinfo {author}
  {\bibfnamefont {M.~B.}\ \bibnamefont {Plenio}}, \bibinfo {author}
  {\bibfnamefont {M.}~\bibnamefont {Huber}}, \bibinfo {author} {\bibfnamefont
  {C.}~\bibnamefont {Roos}}, \bibinfo {author} {\bibfnamefont {R.}~\bibnamefont
  {Blatt}}, \ and\ \bibinfo {author} {\bibfnamefont {B.}~\bibnamefont
  {Lanyon}},\ }\href {\doibase 10.1103/PhysRevX.8.021012} {\bibfield  {journal}
  {\bibinfo  {journal} {Phys. Rev. X}\ }\textbf {\bibinfo {volume} {8}},\
  \bibinfo {pages} {021012} (\bibinfo {year} {2018})}\BibitemShut {NoStop}%
\bibitem [{\citenamefont {Stick}\ \emph {et~al.}(2006)\citenamefont {Stick},
  \citenamefont {Hensinger}, \citenamefont {Olmschenk}, \citenamefont {Madsen},
  \citenamefont {Schwab},\ and\ \citenamefont {Monroe}}]{stick_ion_2006}%
  \BibitemOpen
  \bibfield  {author} {\bibinfo {author} {\bibfnamefont {D.}~\bibnamefont
  {Stick}}, \bibinfo {author} {\bibfnamefont {W.~K.}\ \bibnamefont
  {Hensinger}}, \bibinfo {author} {\bibfnamefont {S.}~\bibnamefont
  {Olmschenk}}, \bibinfo {author} {\bibfnamefont {M.~J.}\ \bibnamefont
  {Madsen}}, \bibinfo {author} {\bibfnamefont {K.}~\bibnamefont {Schwab}}, \
  and\ \bibinfo {author} {\bibfnamefont {C.}~\bibnamefont {Monroe}},\ }\href
  {\doibase 10.1038/nphys171} {\bibfield  {journal} {\bibinfo  {journal} {Nat.
  Phys.}\ }\textbf {\bibinfo {volume} {2}},\ \bibinfo {pages} {36} (\bibinfo
  {year} {2006})}\BibitemShut {NoStop}%
\bibitem [{\citenamefont {Kielpinski}\ \emph {et~al.}(2002)\citenamefont
  {Kielpinski}, \citenamefont {Monroe},\ and\ \citenamefont
  {Wineland}}]{kielpinski_architecture_2002}%
  \BibitemOpen
  \bibfield  {author} {\bibinfo {author} {\bibfnamefont {D.}~\bibnamefont
  {Kielpinski}}, \bibinfo {author} {\bibfnamefont {C.}~\bibnamefont {Monroe}},
  \ and\ \bibinfo {author} {\bibfnamefont {D.~J.}\ \bibnamefont {Wineland}},\
  }\href {\doibase 10.1038/nature00784} {\bibfield  {journal} {\bibinfo
  {journal} {Nature}\ }\textbf {\bibinfo {volume} {417}},\ \bibinfo {pages}
  {709} (\bibinfo {year} {2002})}\BibitemShut {NoStop}%
\bibitem [{\citenamefont {Monroe}\ and\ \citenamefont
  {Kim}(2013)}]{monroe_scaling_2013}%
  \BibitemOpen
  \bibfield  {author} {\bibinfo {author} {\bibfnamefont {C.}~\bibnamefont
  {Monroe}}\ and\ \bibinfo {author} {\bibfnamefont {J.}~\bibnamefont {Kim}},\
  }\href {\doibase 10.1126/science.1231298} {\bibfield  {journal} {\bibinfo
  {journal} {Science}\ }\textbf {\bibinfo {volume} {339}},\ \bibinfo {pages}
  {1164} (\bibinfo {year} {2013})}\BibitemShut {NoStop}%
\bibitem [{\citenamefont {Lekitsch}\ \emph {et~al.}(2017)\citenamefont
  {Lekitsch}, \citenamefont {Weidt}, \citenamefont {Fowler}, \citenamefont
  {Mølmer}, \citenamefont {Devitt}, \citenamefont {Wunderlich},\ and\
  \citenamefont {Hensinger}}]{lekitsch_blueprint_2017}%
  \BibitemOpen
  \bibfield  {author} {\bibinfo {author} {\bibfnamefont {B.}~\bibnamefont
  {Lekitsch}}, \bibinfo {author} {\bibfnamefont {S.}~\bibnamefont {Weidt}},
  \bibinfo {author} {\bibfnamefont {A.~G.}\ \bibnamefont {Fowler}}, \bibinfo
  {author} {\bibfnamefont {K.}~\bibnamefont {Mølmer}}, \bibinfo {author}
  {\bibfnamefont {S.~J.}\ \bibnamefont {Devitt}}, \bibinfo {author}
  {\bibfnamefont {C.}~\bibnamefont {Wunderlich}}, \ and\ \bibinfo {author}
  {\bibfnamefont {W.~K.}\ \bibnamefont {Hensinger}},\ }\href {\doibase
  10.1126/sciadv.1601540} {\bibfield  {journal} {\bibinfo  {journal} {Science
  Advances}\ }\textbf {\bibinfo {volume} {3}},\ \bibinfo {pages} {e1601540}
  (\bibinfo {year} {2017})}\BibitemShut {NoStop}%
\bibitem [{\citenamefont {Wrachtrup}\ \emph {et~al.}(2001)\citenamefont
  {Wrachtrup}, \citenamefont {Kilin},\ and\ \citenamefont
  {Nizovtsev}}]{wrachtrup_quantum_2001}%
  \BibitemOpen
  \bibfield  {author} {\bibinfo {author} {\bibfnamefont {J.}~\bibnamefont
  {Wrachtrup}}, \bibinfo {author} {\bibfnamefont {S.~Y.}\ \bibnamefont
  {Kilin}}, \ and\ \bibinfo {author} {\bibfnamefont {A.~P.}\ \bibnamefont
  {Nizovtsev}},\ }\href {\doibase 10.1134/1.1405224} {\bibfield  {journal}
  {\bibinfo  {journal} {Opt. Spectrosc.}\ }\textbf {\bibinfo {volume} {91}},\
  \bibinfo {pages} {429} (\bibinfo {year} {2001})}\BibitemShut {NoStop}%
\bibitem [{\citenamefont {Maurer}\ \emph {et~al.}(2012)\citenamefont {Maurer},
  \citenamefont {Kucsko}, \citenamefont {Latta}, \citenamefont {Jiang},
  \citenamefont {Yao}, \citenamefont {Bennett}, \citenamefont {Pastawski},
  \citenamefont {Hunger}, \citenamefont {Chisholm}, \citenamefont {Markham},
  \citenamefont {Twitchen}, \citenamefont {Cirac},\ and\ \citenamefont
  {Lukin}}]{maurer_room-temperature_2012}%
  \BibitemOpen
  \bibfield  {author} {\bibinfo {author} {\bibfnamefont {P.~C.}\ \bibnamefont
  {Maurer}}, \bibinfo {author} {\bibfnamefont {G.}~\bibnamefont {Kucsko}},
  \bibinfo {author} {\bibfnamefont {C.}~\bibnamefont {Latta}}, \bibinfo
  {author} {\bibfnamefont {L.}~\bibnamefont {Jiang}}, \bibinfo {author}
  {\bibfnamefont {N.~Y.}\ \bibnamefont {Yao}}, \bibinfo {author} {\bibfnamefont
  {S.~D.}\ \bibnamefont {Bennett}}, \bibinfo {author} {\bibfnamefont
  {F.}~\bibnamefont {Pastawski}}, \bibinfo {author} {\bibfnamefont
  {D.}~\bibnamefont {Hunger}}, \bibinfo {author} {\bibfnamefont
  {N.}~\bibnamefont {Chisholm}}, \bibinfo {author} {\bibfnamefont
  {M.}~\bibnamefont {Markham}}, \bibinfo {author} {\bibfnamefont {D.~J.}\
  \bibnamefont {Twitchen}}, \bibinfo {author} {\bibfnamefont {J.~I.}\
  \bibnamefont {Cirac}}, \ and\ \bibinfo {author} {\bibfnamefont {M.~D.}\
  \bibnamefont {Lukin}},\ }\href {\doibase 10.1126/science.1220513} {\bibfield
  {journal} {\bibinfo  {journal} {Science}\ }\textbf {\bibinfo {volume}
  {336}},\ \bibinfo {pages} {1283} (\bibinfo {year} {2012})}\BibitemShut
  {NoStop}%
\bibitem [{\citenamefont {Doherty}\ \emph {et~al.}(2013)\citenamefont
  {Doherty}, \citenamefont {Manson}, \citenamefont {Delaney}, \citenamefont
  {Jelezko}, \citenamefont {Wrachtrup},\ and\ \citenamefont
  {Hollenberg}}]{doherty_nitrogen-vacancy_2013}%
  \BibitemOpen
  \bibfield  {author} {\bibinfo {author} {\bibfnamefont {M.~W.}\ \bibnamefont
  {Doherty}}, \bibinfo {author} {\bibfnamefont {N.~B.}\ \bibnamefont {Manson}},
  \bibinfo {author} {\bibfnamefont {P.}~\bibnamefont {Delaney}}, \bibinfo
  {author} {\bibfnamefont {F.}~\bibnamefont {Jelezko}}, \bibinfo {author}
  {\bibfnamefont {J.}~\bibnamefont {Wrachtrup}}, \ and\ \bibinfo {author}
  {\bibfnamefont {L.~C.~L.}\ \bibnamefont {Hollenberg}},\ }\href {\doibase
  10.1016/j.physrep.2013.02.001} {\bibfield  {journal} {\bibinfo  {journal}
  {Physics Reports}\ }\textbf {\bibinfo {volume} {528}},\ \bibinfo {pages} {1}
  (\bibinfo {year} {2013})}\BibitemShut {NoStop}%
\bibitem [{\citenamefont {Taminiau}\ \emph {et~al.}(2014)\citenamefont
  {Taminiau}, \citenamefont {Cramer}, \citenamefont {Sar}, \citenamefont
  {Dobrovitski},\ and\ \citenamefont {Hanson}}]{taminiau_universal_2014}%
  \BibitemOpen
  \bibfield  {author} {\bibinfo {author} {\bibfnamefont {T.~H.}\ \bibnamefont
  {Taminiau}}, \bibinfo {author} {\bibfnamefont {J.}~\bibnamefont {Cramer}},
  \bibinfo {author} {\bibfnamefont {T.~v.~d.}\ \bibnamefont {Sar}}, \bibinfo
  {author} {\bibfnamefont {V.~V.}\ \bibnamefont {Dobrovitski}}, \ and\ \bibinfo
  {author} {\bibfnamefont {R.}~\bibnamefont {Hanson}},\ }\href {\doibase
  10.1038/nnano.2014.2} {\bibfield  {journal} {\bibinfo  {journal} {Nature
  Nanotechnology}\ }\textbf {\bibinfo {volume} {9}},\ \bibinfo {pages} {171}
  (\bibinfo {year} {2014})}\BibitemShut {NoStop}%
\bibitem [{\citenamefont {Nemoto}\ \emph {et~al.}(2014)\citenamefont {Nemoto},
  \citenamefont {Trupke}, \citenamefont {Devitt}, \citenamefont {Stephens},
  \citenamefont {Scharfenberger}, \citenamefont {Buczak}, \citenamefont
  {Nöbauer}, \citenamefont {Everitt}, \citenamefont {Schmiedmayer},\ and\
  \citenamefont {Munro}}]{nemoto_photonic_2014}%
  \BibitemOpen
  \bibfield  {author} {\bibinfo {author} {\bibfnamefont {K.}~\bibnamefont
  {Nemoto}}, \bibinfo {author} {\bibfnamefont {M.}~\bibnamefont {Trupke}},
  \bibinfo {author} {\bibfnamefont {S.~J.}\ \bibnamefont {Devitt}}, \bibinfo
  {author} {\bibfnamefont {A.~M.}\ \bibnamefont {Stephens}}, \bibinfo {author}
  {\bibfnamefont {B.}~\bibnamefont {Scharfenberger}}, \bibinfo {author}
  {\bibfnamefont {K.}~\bibnamefont {Buczak}}, \bibinfo {author} {\bibfnamefont
  {T.}~\bibnamefont {Nöbauer}}, \bibinfo {author} {\bibfnamefont {M.~S.}\
  \bibnamefont {Everitt}}, \bibinfo {author} {\bibfnamefont {J.}~\bibnamefont
  {Schmiedmayer}}, \ and\ \bibinfo {author} {\bibfnamefont {W.~J.}\
  \bibnamefont {Munro}},\ }\href {\doibase 10.1103/PhysRevX.4.031022}
  {\bibfield  {journal} {\bibinfo  {journal} {Phys. Rev. X}\ }\textbf {\bibinfo
  {volume} {4}},\ \bibinfo {pages} {031022} (\bibinfo {year}
  {2014})}\BibitemShut {NoStop}%
\bibitem [{\citenamefont {Karzig}\ \emph {et~al.}(2017)\citenamefont {Karzig},
  \citenamefont {Knapp}, \citenamefont {Lutchyn}, \citenamefont {Bonderson},
  \citenamefont {Hastings}, \citenamefont {Nayak}, \citenamefont {Alicea},
  \citenamefont {Flensberg}, \citenamefont {Plugge}, \citenamefont {Oreg},
  \citenamefont {Marcus},\ and\ \citenamefont
  {Freedman}}]{karzig_scalable_2017}%
  \BibitemOpen
  \bibfield  {author} {\bibinfo {author} {\bibfnamefont {T.}~\bibnamefont
  {Karzig}}, \bibinfo {author} {\bibfnamefont {C.}~\bibnamefont {Knapp}},
  \bibinfo {author} {\bibfnamefont {R.~M.}\ \bibnamefont {Lutchyn}}, \bibinfo
  {author} {\bibfnamefont {P.}~\bibnamefont {Bonderson}}, \bibinfo {author}
  {\bibfnamefont {M.~B.}\ \bibnamefont {Hastings}}, \bibinfo {author}
  {\bibfnamefont {C.}~\bibnamefont {Nayak}}, \bibinfo {author} {\bibfnamefont
  {J.}~\bibnamefont {Alicea}}, \bibinfo {author} {\bibfnamefont
  {K.}~\bibnamefont {Flensberg}}, \bibinfo {author} {\bibfnamefont
  {S.}~\bibnamefont {Plugge}}, \bibinfo {author} {\bibfnamefont
  {Y.}~\bibnamefont {Oreg}}, \bibinfo {author} {\bibfnamefont {C.~M.}\
  \bibnamefont {Marcus}}, \ and\ \bibinfo {author} {\bibfnamefont {M.~H.}\
  \bibnamefont {Freedman}},\ }\href {\doibase 10.1103/PhysRevB.95.235305}
  {\bibfield  {journal} {\bibinfo  {journal} {Phys. Rev. B}\ }\textbf {\bibinfo
  {volume} {95}},\ \bibinfo {pages} {235305} (\bibinfo {year}
  {2017})}\BibitemShut {NoStop}%
\bibitem [{\citenamefont {Freedman}\ \emph {et~al.}(2003)\citenamefont
  {Freedman}, \citenamefont {Kitaev}, \citenamefont {Larsen},\ and\
  \citenamefont {Wang}}]{freedman_topological_2003}%
  \BibitemOpen
  \bibfield  {author} {\bibinfo {author} {\bibfnamefont {M.}~\bibnamefont
  {Freedman}}, \bibinfo {author} {\bibfnamefont {A.}~\bibnamefont {Kitaev}},
  \bibinfo {author} {\bibfnamefont {M.}~\bibnamefont {Larsen}}, \ and\ \bibinfo
  {author} {\bibfnamefont {Z.}~\bibnamefont {Wang}},\ }\href {\doibase
  10.1090/S0273-0979-02-00964-3} {\bibfield  {journal} {\bibinfo  {journal}
  {Bull. Amer. Math. Soc.}\ }\textbf {\bibinfo {volume} {40}},\ \bibinfo
  {pages} {31} (\bibinfo {year} {2003})}\BibitemShut {NoStop}%
\bibitem [{\citenamefont {Hornibrook}\ \emph {et~al.}(2015)\citenamefont
  {Hornibrook}, \citenamefont {Colless}, \citenamefont {Conway~Lamb},
  \citenamefont {Pauka}, \citenamefont {Lu}, \citenamefont {Gossard},
  \citenamefont {Watson}, \citenamefont {Gardner}, \citenamefont {Fallahi},
  \citenamefont {Manfra},\ and\ \citenamefont
  {Reilly}}]{hornibrook_cryogenic_2015}%
  \BibitemOpen
  \bibfield  {author} {\bibinfo {author} {\bibfnamefont {J.}~\bibnamefont
  {Hornibrook}}, \bibinfo {author} {\bibfnamefont {J.}~\bibnamefont {Colless}},
  \bibinfo {author} {\bibfnamefont {I.}~\bibnamefont {Conway~Lamb}}, \bibinfo
  {author} {\bibfnamefont {S.}~\bibnamefont {Pauka}}, \bibinfo {author}
  {\bibfnamefont {H.}~\bibnamefont {Lu}}, \bibinfo {author} {\bibfnamefont
  {A.}~\bibnamefont {Gossard}}, \bibinfo {author} {\bibfnamefont
  {J.}~\bibnamefont {Watson}}, \bibinfo {author} {\bibfnamefont
  {G.}~\bibnamefont {Gardner}}, \bibinfo {author} {\bibfnamefont
  {S.}~\bibnamefont {Fallahi}}, \bibinfo {author} {\bibfnamefont
  {M.}~\bibnamefont {Manfra}}, \ and\ \bibinfo {author} {\bibfnamefont
  {D.}~\bibnamefont {Reilly}},\ }\href {\doibase
  10.1103/PhysRevApplied.3.024010} {\bibfield  {journal} {\bibinfo  {journal}
  {Phys. Rev. Applied}\ }\textbf {\bibinfo {volume} {3}},\ \bibinfo {pages}
  {024010} (\bibinfo {year} {2015})}\BibitemShut {NoStop}%
\bibitem [{\citenamefont {Asaad}\ \emph {et~al.}(2016)\citenamefont {Asaad},
  \citenamefont {Dickel}, \citenamefont {Langford}, \citenamefont {Poletto},
  \citenamefont {Bruno}, \citenamefont {Rol}, \citenamefont {Deurloo},\ and\
  \citenamefont {DiCarlo}}]{asaad_independent_2016}%
  \BibitemOpen
  \bibfield  {author} {\bibinfo {author} {\bibfnamefont {S.}~\bibnamefont
  {Asaad}}, \bibinfo {author} {\bibfnamefont {C.}~\bibnamefont {Dickel}},
  \bibinfo {author} {\bibfnamefont {N.~K.}\ \bibnamefont {Langford}}, \bibinfo
  {author} {\bibfnamefont {S.}~\bibnamefont {Poletto}}, \bibinfo {author}
  {\bibfnamefont {A.}~\bibnamefont {Bruno}}, \bibinfo {author} {\bibfnamefont
  {M.~A.}\ \bibnamefont {Rol}}, \bibinfo {author} {\bibfnamefont
  {D.}~\bibnamefont {Deurloo}}, \ and\ \bibinfo {author} {\bibfnamefont
  {L.}~\bibnamefont {DiCarlo}},\ }\href {\doibase 10.1038/npjqi.2016.29}
  {\bibfield  {journal} {\bibinfo  {journal} {npj Quantum Information}\
  }\textbf {\bibinfo {volume} {2}},\ \bibinfo {pages} {16029} (\bibinfo {year}
  {2016})}\BibitemShut {NoStop}%
\bibitem [{\citenamefont {Almudever}\ \emph {et~al.}(2017)\citenamefont
  {Almudever}, \citenamefont {Lao}, \citenamefont {Fu}, \citenamefont
  {Khammassi}, \citenamefont {Ashraf}, \citenamefont {Iorga}, \citenamefont
  {Varsamopoulos}, \citenamefont {Eichler}, \citenamefont {Wallraff},
  \citenamefont {Geck}, \citenamefont {Kruth}, \citenamefont {Knoch},
  \citenamefont {Bluhm},\ and\ \citenamefont
  {Bertels}}]{almudever_engineering_2017}%
  \BibitemOpen
  \bibfield  {author} {\bibinfo {author} {\bibfnamefont {C.~G.}\ \bibnamefont
  {Almudever}}, \bibinfo {author} {\bibfnamefont {L.}~\bibnamefont {Lao}},
  \bibinfo {author} {\bibfnamefont {X.}~\bibnamefont {Fu}}, \bibinfo {author}
  {\bibfnamefont {N.}~\bibnamefont {Khammassi}}, \bibinfo {author}
  {\bibfnamefont {I.}~\bibnamefont {Ashraf}}, \bibinfo {author} {\bibfnamefont
  {D.}~\bibnamefont {Iorga}}, \bibinfo {author} {\bibfnamefont
  {S.}~\bibnamefont {Varsamopoulos}}, \bibinfo {author} {\bibfnamefont
  {C.}~\bibnamefont {Eichler}}, \bibinfo {author} {\bibfnamefont
  {A.}~\bibnamefont {Wallraff}}, \bibinfo {author} {\bibfnamefont
  {L.}~\bibnamefont {Geck}}, \bibinfo {author} {\bibfnamefont {A.}~\bibnamefont
  {Kruth}}, \bibinfo {author} {\bibfnamefont {J.}~\bibnamefont {Knoch}},
  \bibinfo {author} {\bibfnamefont {H.}~\bibnamefont {Bluhm}}, \ and\ \bibinfo
  {author} {\bibfnamefont {K.}~\bibnamefont {Bertels}},\ }in\ \href {\doibase
  10.23919/DATE.2017.7927104} {\emph {\bibinfo {booktitle} {Design,
  {Automation} {Test} in {Europe} {Conference} {Exhibition} ({DATE}), 2017}}}\
  (\bibinfo {year} {2017})\ pp.\ \bibinfo {pages} {836--845}\BibitemShut
  {NoStop}%
\bibitem [{\citenamefont {Suchara}\ \emph {et~al.}(2013)\citenamefont
  {Suchara}, \citenamefont {Faruque}, \citenamefont {Lai}, \citenamefont {Paz},
  \citenamefont {Chong},\ and\ \citenamefont
  {Kubiatowicz}}]{suchara_comparing_2013}%
  \BibitemOpen
  \bibfield  {author} {\bibinfo {author} {\bibfnamefont {M.}~\bibnamefont
  {Suchara}}, \bibinfo {author} {\bibfnamefont {A.}~\bibnamefont {Faruque}},
  \bibinfo {author} {\bibfnamefont {C.-Y.}\ \bibnamefont {Lai}}, \bibinfo
  {author} {\bibfnamefont {G.}~\bibnamefont {Paz}}, \bibinfo {author}
  {\bibfnamefont {F.~T.}\ \bibnamefont {Chong}}, \ and\ \bibinfo {author}
  {\bibfnamefont {J.}~\bibnamefont {Kubiatowicz}},\ }\href
  {http://arxiv.org/abs/1312.2316} {\bibfield  {journal} {\bibinfo  {journal}
  {arXiv:1312.2316}\ } (\bibinfo {year} {2013})}\BibitemShut {NoStop}%
\end{thebibliography}%
\end{document}